\documentclass[12pt]{article}
\pdfoutput=1 

\usepackage{jheppub} 
\usepackage[T1]{fontenc} 

\title{\boldmath Construction of consistent interactions in higher derivative Yang-Mills gauge theory with matter fields}

\author[]{J.L. Dai,\note{Corresponding author.}}

\affiliation[a]{Department of Physics, Zhejiang University, Hangzhou, 310027, P. R. China}
\affiliation[b]{Center of Mathematical Science, Zhejiang University, Hangzhou, 310027, P. R. China}

\emailAdd{daijlxy@126.com}

\abstract{We show the consistent interactions in the generalized electrodynamics gauge theory with higher derivative matter fields by means of the order reduction method. We deduce the BRST deformations in the reduced Lagrangian and using the equations of motion of the auxiliary fields in the antighost number zero part in the resulting deformed action, we are able to obtain the consistent coupling terms added into the original Lagrangian density which are compatible with the deformation master equations. We emphasize that the order of deformations is truncated at four and the corresponding higher-order deformations are equal to zero precisely. Moreover, the local Abelian gauge symmetry turns out to be non-Abelian after the deformation procedure.}

\begin{document}
\maketitle
\flushbottom

\section{Introduction}

Higher derivatives [1,2] appear frequently in effective field theories through the higher order operators and almost any effective theory obtained by integrating out some degrees of freedom of the underlying theory usually contains higher derivatives. The first model of a higher-order derivative field theory is a generalization of the electromagnetic field proposed in the works of Podolsky [3] who suggested to modify the Maxwell-Lorentz theory in order to avoid divergences such as the electron self-energy and the vacuum polarization current which were caused by the singularity in classical electrodynamics. In such generalized electrodynamics gauge theory, the action is modified by a second-order derivative of the gauge field term and therefore, the equations of motion turn out to be fourth-order partial differential equations but are still linear in the fields [4,5]. In the context of gravity, higher-derivative modifications have been studied extensively since they greatly improve the renormalization properties of field theories [6-8] in four dimensions. More importantly, the existence of these higher order terms will alter the effective potential and phase transitions of scalar fields in curved spacetime and produce a profound influence on the research of astrophysical and cosmological behaviors [9,10]. For instance, the higher derivative scalar fields play very important
role in the study of inflation of the universe which thus can be selected as candidates of dark matter of the universe [11]. Higher derivative scalar fields are also present in supersymmetric field theories and have been studied very intensively in the literature [12,13]. Furthermore, in the research of effective action for the trace anomaly in conformal field theory [14], the higher derivative fermionic field theories were considered as a dynamical mechanism for fermionic mass generation. In the higher derivative gravity [15], although the Faddeev-Popov compensating fields are not fermionic fields, they are anticommuting ones and therefore it is inevitable to study the higher derivative fermionic systems. Besides, compared to the experimental results of the $W_{\pm},Z$ and $\gamma$ self-interactions and of the Higgs sector in standard model, it is possible to consider arbitrary interaction vertices of massive vector and scalar fields in effective Lagrangians to explore the deviations of electroweak interactions. Such idea can be achieved by adding higher order derivative terms in standard Yang-Mills vector-boson self-interactions models. Therefore, the investigation of effective Lagrangians with higher order derivatives is very significant from the theoretical and phenomenological point of view.

Recent works have applied the path integral formalism and the BFV approaches to explore the quantization of the generalized electrodynamic systems [16,17]. In addition, such higher derivative Lagrangian also has been investigated by means of the order reduction method in Faddeev-Jackiw quantization scheme [18]. Like the usual theories, the higher derivative field theories may be endowed with gauge symmetry which implies that these systems possess more degrees of freedom than the physical ones and hence we might encounter constrains of dynamic variables [19,20]. The most powerful tool to treat these constraints in gauge system is the BRST transformation developed by Becchi, Rouet, Stora and Tyutin [21-23] through introducing a set of anticommutative variables associated to every constraint and this idea provides an algebraic technique to deal with the renormalization difficulties and explain anomalous phenomena in gauge theories [24,25]. The central role in this formulation is the nilpotent BRST generator which generally can be divided into two parts, namely the Koszul-Tate differential derivative and the exterior longitudinal derivative along the gauge orbits on the constraint surface proposed by Henneaux et al [26,27]. Under such decomposition, the inequivalent infinitesimal local gaugings corresponding to BRST cohomology classes in ghost number zero are characterized in the space of local functionals. Also an isomorphism has been established between these cohomology classes and the algebra of observables, or more concretely the gauge invariant functionals in the original constraint theory [28-30]. While at the negative ghost number, the BRST cohomology is identified with the "Characteristic cohomology" as pointed out in [28,31] that can be interpreted as the generalization of the usual conserved currents in classical field theories. Moreover, in the consideration of the BRST cohomology with positive ghost number, we are certain to construct the consistent deformations in a gauge invariant action within the framework of antifield formulation [32,33].

More specifically, as is well known, the Batalin-Vilkovisky (BV) formalism [34-36] is a natural extension of the BRST approach by doubling a collection of antifields with opposite statistics compared to the classical ones. The core of the BV formalism is the anticanonical bracket on the space of the local functionals of the whole fields/antifields together with the existence of the master action $S_{0}$ which fulfills the so-called master equation $(S_{0},S_{0})=0$ and the original action is viewed as its boundary condition [37,38]. The extended BRST transformation associated to the fields/antifields $(\phi^{A},\phi^{*}_{A})$ can in general be read directly from this antibracket with the master action $S_{0}$ [26]. When restricted to the ordinary gauge theory, it is believed that we will recover the results of BRST formulation mentioned above. Furthermore, through solving the deformations of the master equation, we are able to derive the consistent interactions among the gauge fields with the aid of the extended BRST cohomology in the antifield formalism [39,40]. In principle, such BRST deformation procedure can be employed to higher-order deformations where obstructions will arise naturally and the number of the gauge symmetries is kept after the deformations. In the pure gauge theory, it is possible to add gauge invariant terms to the Lagrangian without changing the gauge symmetries [41-43], while coupled to matter fields, the required consistent deformations can be deduced from the conservation of the currents resulting from the global invariance of the original gauge-matter system [44,45]. Besides, the deformation procedure will alter the form of the gauge transformations of the matter fields and the deformation parameter can be regarded as the couplings constant among fields. Plenty of research has been made on the area of the deformations of the master action in various contexts of gauge theories through the analysis of BRST cohomology by Bizdadea et al [46-49] and the recent developments of BRST consistent deformations can be consulted in [50-52].

In this work, we mainly intend to construct the consistent interactions in the generalized electrodynamics gauge theory with higher derivative matter fields including real scalar, complex scalar and Dirac spinor fields. At first, we reduce this system to an equivalent lower-order model with the help of the extra auxiliary gauge and matter fields which is more manageable due to the simplicity arising from lower-order derivative terms. Then by a detailed analysis of the BRST cohomology of the reduced system, we are capable of deriving the first- and second-order deformations from the deformation master equations and it can be proved that the corresponding higher-order deformation terms vanish completely. After the deformation procedure, we extract the antighost number zero part in the deformed master action and taking advantage of the equations of motion of the auxiliary fields, one can convert this resulting Lagrangian into an equivalent action containing of interactions expressed in terms of the classical gauge and matter fields. Comparing to the original higher derivative system, we conclude that these extra interaction terms are the consistent deformations we search for and indeed they satisfy the deformation master equations but with tedious checks.

This paper is organized as follows. In section 2, we simply study the order reduced system of the generalized electrodynamics theory with higher derivative real scalar fields and review the standard BRST deformations of the general irreducible gauge theory in the BV formalism. Based on these results, we show the cohomological derivation of the consistent deformations in this free higher derivative gauge theory. Also, we calculate the path integral of the higher derivative systems in the BRST-BV quantization scheme by choosing appropriate gauge-fixing fermions and simply discuss the quantum master equation as the generalization of classical ones.  Section 3,4 are devoted to the construction of consistent interactions in the situations of complex scalar fields and Dirac spinor fields parallel to the discussion in the real scalar case. The final section of this paper is for conclusion and further works.

\section{Massless real scalar fields}
\subsection{BRST deformation}
Let us consider the free generalized electrodynamics gauge theory with higher derivative matter fields described by a set of $N$ gauge fields $A_{\mu}^{a}$ for $a=1,...,N$ and a collection of real scalar fields $\phi^{i}$ for $i=1,2,...,M$ in (3+1)-dimensional spacetime with metric $g_{\mu\nu}=\mathrm{diag}(-1,-1,-1,1)$ as follows
\begin{equation}
\begin{aligned}
\mathcal{L}(A_{\mu}^{a},\phi^{i})=-\frac{1}{4}F_{\mu\nu}^{a}F^{\mu\nu}_{a}+\frac{m^{2}}{2}\partial_{\mu}F^{\mu\nu}_{a}\partial^{\lambda}F_{\lambda\nu}^{a}+\frac{1}{2}G_{ij}(\partial_{\mu}\phi^{i}\partial^{\mu}\phi^{j}-\frac{1}{\alpha}\square\phi^{i}\square\phi^{j})
\end{aligned}
\end{equation}
here $F_{\mu\nu}^{a}=\partial_{\mu}A_{\nu}^{a}-\partial_{\nu}A_{\mu}^{a}$ is the usual electromagnetic field strength tensor and we use the metric $g_{\mu\nu}$ raises and lows the indices, $m,\alpha$ are some constants and $G_{ij}$ is an invertible symmetric constant matrix. It is obvious that since the presence of the real scalar fields, the Lagrangian density $\mathcal{L}$ is invariant under the following Abelian local gauge transformations with rigid symmetries
\begin{equation}
\begin{aligned}
A^{a}_{\mu}\rightarrow A^{'a}_{\mu}=A^{a}_{\mu}+\partial_{\mu}\lambda^{a},\quad \quad \phi^{i} \rightarrow \phi^{'i}=\phi^{i}
\end{aligned}
\end{equation}
here $\lambda^{a}$ are arbitrary functions. By variation with respect to the dynamic variables $A_{\mu}^{a},\phi^{i}$, the fourth-order equations of motion for gauge fields and scalar fields are given by
\begin{equation}
\begin{aligned}
(1+m^{2}\square)(g^{\mu\nu}\square-\partial^{\mu}\partial^{\nu})A_{\nu}^{a}=0,\quad \quad \quad   G_{ij}\square(\square+\alpha)\phi^{j}=0
\end{aligned}
\end{equation}

It is well known that the presence of higher derivative terms in any non-degenerate Lagrangian will give rise to unbounded trajectories
in the classical regime and to loss of unitarity at the quantum level. Such so-called Ostrogradsky instability yields the Hamiltonian of the system is not bounded from below, or in other words the energy can be lowered without any bound by increasing the momentum to large negative values [53]. Unfortunately, this notorious phenomenon generally can not be cured by trying to do any alternative canonical transformations. Recently, a class of higher derivative theories of derived type was discussed in [54-56]. In these models, the “derived” means that the wave operator determining the equations of motion of the theory can be formulated through a polynomial of arbitrary finite order in another lower order differential operator. This lower order operator is supposed to be of the first or second order and is termed as primary wave operator. At the free level, a symmetry of the higher derivative system is a linear operator which is interchangeable with the primary wave operator in certain condition. The existence of such symmetry will lead to $n$ independent higher order symmetries of the field equations if the order of the characteristic polynomial of the wave operator is $n$. When combining these generators of higher order symmetries together, we will obtain a series of $n$-parametric derived symmetries of the derived system which is connected to a series of independent conserved quantities by the extension of Noether's theorem. In particular, as one of the most simplest symmetry of the primary wave operator, the spacetime translation invariance will produce a series of conserved second-rank tensors including the standard canonical energy-momentum tensors and the others are different independent integrals of motion. Although the canonical energy is unbounded in higher derivative system, these series of conserved tensors can be bounded and thus the theory is stable in classical regime, which also persists at quantum level [54]. Such analysis of stability in higher derivative dynamics has been employed to the extended Chern-Simons theory coupled to a charged scalar field in [57-59] and the related results can be directly applied to the generalized electrodynamics. Therefore, the higher derivative Yang-Mills systems with matter fields we investigate are stable both in free and coupling cases and by this reason, they are considered as physically acceptable models.

In order to deal with the higher-order derivative terms in (2.1), we introduce a set of auxiliary fields $B^{a}_{\mu}, Z^{i}$ to reduce the order of the original Lagrangian density [18]
\begin{equation}
\begin{aligned}
\mathcal{\tilde{L}}=&-\frac{1}{4}F_{\mu\nu}^{a}F^{\mu\nu}_{a}-\frac{m^{2}}{2}B^{a}_{\mu}B^{\mu }_{a}+m^{2}\partial_{\mu}B_{\nu}^{a}F^{\mu\nu}_{a}+\frac{1}{2}G_{ij}(\partial_{\mu}\phi^{i}\partial^{\mu}Z^{j}\\
&+\frac{1}{4}\alpha\phi^{i}\phi^{j}+\frac{1}{4}\alpha Z^{i}Z^{j}-\frac{1}{2}\alpha\phi^{i} Z^{j})
\end{aligned}
\end{equation}
which  immediately gives us the coupled equations of motion
\begin{equation}
\begin{aligned}
&(g^{\mu\nu}\square-\partial^{\mu}\partial^{\nu})A_{\nu}^{a}=m^{2}(g^{\mu\nu}\square-\partial^{\mu}\partial^{\nu})B_{\nu}^{a},\quad \quad \quad (g^{\mu\nu}\square-\partial^{\mu}\partial^{\nu})A_{\nu}^{a}=-B^{\mu a},\\
&G_{ij}(1+2\frac{\square}{\alpha})\phi^{i}=G_{ij}Z^{i},\quad \quad \quad \quad \quad\quad \quad \quad \quad \quad  \quad  G_{ij}(1+2\frac{\square}{\alpha})Z^{i}=G_{ij}\phi^{i}
\end{aligned}
\end{equation}
it is convenient to examine that the dynamical equations (2.3) can be retrieved after simple algebraic manipulations from these coupled equations and therefore the reduced Lagrangian is equivalent to the original one. Noting that the invariance of the Lagrangian (2.4) is guaranteed by the following local gauge transformations
\begin{equation}
\begin{aligned}
\Delta_{\lambda} A_{\mu}^{a}=\partial_{\mu}\lambda^{a},\quad \quad \Delta_{\lambda} B_{\mu}^{a}=0,\quad  \quad\Delta_{\lambda}\phi^{i}=0,\quad \quad \Delta_{\lambda} Z^{i}=0
\end{aligned}
\end{equation}

Within the standard BV framework, during our classical treatment discussed above, we should add the ghost fields $\eta^{a}$ due to the emergence of the irreducible generating set of the gauge symmetries and let us denote the extended set of fields by $\phi^{A}=(A^{a}_{\mu},B^{a}_{\mu},\phi^{i},Z^{i},\eta^{a})$. Also to get a bigger conceptual picture, a collection of antifields $\phi^{*}_{A}=(A_{a}^{*\mu},B_{a}^{*\mu},\phi^{*}_{i},Z^{*}_{i},\eta^{*}_{a})$ with opposite statistics compared to their partners are needed for the Koszul-Tate resolution of the equations of motion. Then it is customary to introduce the Grassmann parities, antighost, pureghost and ghost numbers both for the fields and antifields as follows [28,45]
\begin{equation}
\begin{aligned}
&\epsilon(A_{\mu}^{a},B_{\mu}^{a},\phi^{i},Z^{i})=0,\quad\quad\epsilon(A_{a}^{*\mu},B_{a}^{*\mu},\phi^{*}_{i},Z^{*}_{i})=1,\quad\quad\epsilon(\eta^{a})=1,\quad \quad \epsilon(\eta^{*}_{a})=0,\\
&\mathrm{agh}(A_{\mu}^{a},B_{\mu}^{a},\phi^{i},Z^{i})=0,\quad \mathrm{agh}(A_{a}^{*\mu},B_{a}^{*\mu},\phi^{*}_{i},Z^{*}_{i})=1,\quad  \mathrm{agh}(\eta^{a})=0,\quad \mathrm{agh}(\eta^{*}_{a})=2,\\
&\mathrm{pgh}(A_{\mu}^{a},B_{\mu}^{a},\phi^{i},Z^{i})=0,\quad \mathrm{pgh}(A_{a}^{*\mu},B_{a}^{*\mu},\phi^{*}_{i},Z^{*}_{i})=0,\quad  \mathrm{pgh}(\eta^{a})=1,\quad \mathrm{pgh}(\eta^{*}_{a})=0,\\
&\mathrm{gh}(A_{\mu}^{a},B_{\mu}^{a},\phi^{i},Z^{i})=0,\quad \quad \mathrm{gh}(A_{a}^{*\mu},B_{a}^{*\mu},\phi^{*}_{i},Z^{*}_{i})=-1,\quad \mathrm{gh}(\eta^{a})=1,\quad \mathrm{gh}(\eta^{*}_{a})=-2\\
\end{aligned}
\end{equation}

The local functionals of the whole fields $(\phi^{A},\phi^{*}_{A})$ constitute an odd symplectic manifold which is endowed with an odd Poisson bracket or called antibracket, that is, for any two local functionals $F(\phi^{A},\phi^{*}_{A}),G(\phi^{A},\phi^{*}_{A})$, the antibracket is defined by [26]
\begin{equation}
\begin{aligned}
(F,G)=\int_{M}(\frac{\partial_{r}F}{\partial\phi^{A}}\frac{\partial_{l}G}{\partial\phi_{A}^{*}}-\frac{\partial_{r}G}{\partial\phi^{A}}\frac{\partial_{l}F}{\partial\phi_{A}^{*}})d^{4}x
\end{aligned}
\end{equation}
here $l,r$ refer to the left and right derivative respectively and the summation over $A$ is understand. This antibracket satisfies graded commutation, distribution and Jacobi relations while the fields and antifields under this operation behave as coordinates and momenta. In particular, the antibracket has ghost number 1.

On the other hand, it is well known that in the BV formalism of the general irreducible gauge theory, the master action of the above gauge system can be constructed from (2.4) as its boundary condition, then add higher order terms by associating every antifield an irreducible generating set of gauge transformations that the gauge parameters are replaced by ghosts. In the current study, by means of the local gauge transformation (2.6), the minimal solution of the master action $S_{0}\left[A^{a}_{\mu},B^{a}_{\mu},\phi^{i},Z^{i},A^{*\mu}_{a},\eta^{a}\right]$ admits the form of
\begin{equation}
\begin{aligned}
S_{0}=\int d^{4}x(&-\frac{1}{4}F_{\mu\nu}^{a}F^{\mu\nu}_{a}-\frac{m^{2}}{2}B^{a}_{\mu}B^{\mu }_{a}+m^{2}\partial_{\mu}B_{\nu}^{a}F^{\mu\nu}_{a}+\frac{1}{2}G_{ij}(\partial_{\mu}\phi^{i}\partial^{\mu}Z^{j}\\
&+\frac{1}{4}\alpha\phi^{i}\phi^{j}+\frac{1}{4}\alpha Z^{i}Z^{j}-\frac{1}{2}\alpha\phi^{i} Z^{j})+A^{*\mu}_{a}\partial_{\mu}\eta^{a})
\end{aligned}
\end{equation}
which is a functional of ghost number $0$ and the choice of the master action $S_{0}$ is uniquely defined up to anticanonical transformations as the BRST charge does.

At this point we stress that the BRST transformation associated to the fields/antifields $(\phi^{A},\phi^{*}_{A})$ can be determined from this antibracket with the master action $S_{0}$ through the manner as
\begin{equation}
\begin{aligned}
s=(S_{0},\cdot)
\end{aligned}
\end{equation}
as explained in [26,27], the BRST generator $s=\delta+\gamma$ of the general irreducible gauge theory can be furthre decomposed into two parts called Koszul-Tate differential $\delta$ and the exterior longitudinal derivative $\gamma$. In the present situation, they act on the generators of the BRST complex in the following way
\begin{equation}
\begin{aligned}
&\delta A^{a}_{\mu}=\delta B^{a}_{\mu}=\gamma B^{a}_{\mu}=\delta\eta^{a}=\gamma\eta^{a}=0,\quad \quad \quad \gamma A^{a}_{\mu}=\partial_{\mu}\eta^{a},\\
&\delta A^{*\mu}_{a}=\partial_{\nu}F^{\mu\nu}_{a}+m^{2}(\square B^{\mu}_{a}-\partial^{\mu}\partial_{\nu}B^{\nu}_{a}),\quad \quad \quad \delta B^{*\mu}_{a}=m^{2}(B^{\mu}_{a}+\partial_{\nu}F^{\nu\mu}_{a}),\\
&\gamma A^{*\mu}_{a}=\gamma B^{*\mu}_{a}=\gamma\eta^{*}_{a}=0,\quad \quad \quad \quad \quad \quad \quad \quad \delta\eta^{*}_{a}=-\partial_{\mu}A^{*\mu}_{a}
\end{aligned}
\end{equation}
together with
\begin{equation}
\begin{aligned}
&\delta\phi^{i}=\gamma\phi^{i}=\gamma\phi^{*}_{i}=0,\quad \quad \delta\phi^{*}_{i}=\frac{1}{2}G_{ij}(\square Z^{j}-\frac{1}{2}\alpha\phi^{j}+\frac{1}{2}\alpha Z^{j}),\\
&\delta Z^{i}=\gamma Z^{i}=\gamma Z^{*}_{i}=0,\quad \quad \delta Z^{*}_{i}=\frac{1}{2}G_{ij}(\square \phi^{j}-\frac{1}{2}\alpha Z^{j}+\frac{1}{2}\alpha \phi^{j})
\end{aligned}
\end{equation}

At the quantum level, to calculate the path integral of system (2.9), a gauge condition is necessary to remove redundant degrees of freedom and for the purpose of implementing the covariant Lorentz gauge,one needs to add a non-minimal term to (2.9) which has no influence on the solution of classical master equation,namely
\begin{equation}
\begin{aligned}
\int d^{4}x\bar{\eta}^{*a}\lambda_{a}
\end{aligned}
\end{equation}
 where $\bar{\eta}_{a}$ are the antighosts of ghost number minus one, $\lambda_{a}$ are the auxiliary field and $\bar{\eta}^{*a},\lambda^{*a}$ are the corresponding antifields. In this way, the total solution of classical master equation is replaced by [27]
\begin{equation}
\begin{aligned}
S_{0}=\int d^{4}x(\mathcal{\tilde{L}}+A^{*\mu}_{a}\partial_{\mu}\eta^{a}+\bar{\eta}^{*a}\lambda_{a})
\end{aligned}
\end{equation}
now choosing a family of gauge-fixing fermions
\begin{equation}
\begin{aligned}
\Psi=\int d^{3}x\bar{\eta}_{a}(\partial^{\mu}A_{\mu}^{a}-\frac{\alpha}{2}\lambda_{a})
\end{aligned}
\end{equation}
here $\alpha$ is an arbitrary constant, particularly the transition amplitude of the original gauge theory is independent of $\alpha$.
If one inserts (2.15) into the solution $S_{0}\left[\phi^{A},\phi^{*}_{A}\right]$ of the master equation, one gets
the "gauge-fixed action" $S_{\Psi}$
\begin{equation}
\begin{aligned}
S_{\Psi}=S_{0}\left[\phi^{A},\phi^{*}_{A}=\frac{\delta\Psi}{\delta\phi^{A}}\right]
\end{aligned}
\end{equation}
and from (2.15) we have the following identification
\begin{equation}
\begin{aligned}
\bar{\eta}^{*a}=-(\partial^{\mu}A_{\mu}^{a}-\frac{\alpha}{2}\lambda^{a}),\quad A^{\ast\mu}_{a}=\partial^{\mu}\bar{\eta}_{a}
\end{aligned}
\end{equation}
which gives rise to the effective action
\begin{equation}
\begin{aligned}
S_{\Psi}=\int d^{4}x\left[\mathcal{\tilde{L}}-(\partial^{\mu}A_{\mu}^{a}-\frac{\alpha}{2}\lambda^{a})\lambda^{a}-\bar{\eta}_{a}\square\eta^{a}\right]
\end{aligned}
\end{equation}
here $\square $ is the usual D'Alembert operator.  After integration over the auxiliary fields $\lambda_{a}$, when $\alpha\neq0$, the path integral becomes
 \begin{equation}
\begin{aligned}
Z=\int\left[DA_{\mu}^{a}\right]\left[DB_{\mu}^{a}\right]\left[D\phi^{i}\right]\left[DZ^{i}\right]\left[D\eta^{a}\right]\left[D\bar{\eta}_{a}\right]\mathrm{exp}\frac{i}{\hbar}\int d^{4}x\left[\mathcal{\tilde{L}}-\frac{1}{2\alpha}(\partial^{\mu}A_{\mu}^{a})^{2}-\bar{\eta}_{a}\square\eta^{a}\right]
\end{aligned}
\end{equation}
in the above expression, when $\alpha=1$  we simply obtain the Feynman gauge, while the choice of $\alpha=0$ in (2.18) results in the Landau gauge.

Having at hand a solution of the classical master equation, it is of interest to deform this action to construct the consistent interactions as follows
\begin{equation}
\begin{aligned}
S=S_{0}+gS_{1}+g^{2}S_{2}+......
\end{aligned}
\end{equation}
substitution of this deformed quantity into the master equation $(S,S)=0$ will lead to a set of deformation master equations from different orders [43]
\begin{equation}
\begin{aligned}
&1:(S_{0},S_{0})=0,\\
&g^{1}:2(S_{0},S_{1})=0,\\
&g^{2}:2(S_{0},S_{2})+(S_{1},S_{1})=0,\\
&g^{3}:(S_{0},S_{3})+(S_{1},S_{2})=0,\\
&......
\end{aligned}
\end{equation}
thus, the burden of a construction of consistent interactions in the master action is to solve all these deformation equations at arbitrary order $g$, and the deformation parameter $g$ is thought of as representing couplings among the gauge and matter fields.

More generally, at the quantum level, we assume the transition amplitude takes the integral form of $\int \mathrm{exp}(i/\hbar)W$, here $W$ is a local functional of the fields/antifields which differs from the classical master action by terms of higher order expansions in $\hbar$
\begin{equation}
\begin{aligned}
W=S+\hbar M_{1}+\hbar^{2} M_{2}+......, \quad \quad \quad \quad \mathrm{gh}W=0
\end{aligned}
\end{equation}
then the gauge-fixed integral is given by
\begin{equation}
\begin{aligned}
Z=\int \left[D\phi^{A}\right]\mathrm{exp}(i/\hbar)W(\phi^{A},\phi^{*}_{A}=\frac{\delta\Psi}{\delta\phi^{A}})
\end{aligned}
\end{equation}
for such path integral to be independent of the gauge-fixing fermion $\Psi$, $\mathrm{exp}(i/\hbar)W$ should be $\Delta$-closed [27], that is to say
\begin{equation}
\begin{aligned}
\Delta\mathrm{exp}(i/\hbar)W=0, \quad \quad  \Delta=(-1)^{\epsilon(A)+1}\frac{\delta^{R}}{\delta\phi^{A}}\frac{\delta^{R}}{\delta\phi^{*}_{A}},\quad \quad \Delta^{2}=0
\end{aligned}
\end{equation}
expanding the above expression in terms of $\hbar$, we are thus led to the equivalent condition
\begin{equation}
\begin{aligned}
i\hbar\Delta W-\frac{1}{2}(W,W)=0
\end{aligned}
\end{equation}
this is the so-called quantum master equation in  Batalin-Vilkovisky formalism and to the zeroth order in $\hbar$, the quantum master equation reduces to the classical master equation $(S,S)=0$. In this paper, we mainly focus on the study of the consistent deformations derived from the classical master equation and do not intend to learn such quantum master equation.

\subsection{First-order deformation}
In this section, we shall confine our attention to a derivation of the solution of the first-order deformation term $S_{1}=\int d^{4}x a_{1}$ in (2.21) which is satisfied by the equation in the simple form of
\begin{equation}
\begin{aligned}
0=sS_{1}=\int d^{4}x sa_{1}
\end{aligned}
\end{equation}
using the decomposition of the BRST operator $s$, it is instructive to rewrite it in the local form
\begin{equation}
\begin{aligned}
sa_{1}=\delta a_{1}+\gamma a_{1} =\partial_{\mu}V^{\mu}_{1}
\end{aligned}
\end{equation}
here we assume $a_{1} $ fulfill $\mathrm{gh}(a_{1} )=\epsilon(a_{1} )=0$. To search for the solution of (2.27), the basic idea is to formulate the $a_{1} $  according to the antighost number
\begin{equation}
\begin{aligned}
a_{1}=a_{1}^{(0)} +a_{1}^{(1)} +...+a_{1}^{(I)}
\end{aligned}
\end{equation}
here $\mathrm{agh}(a_{1}^{(i)})=i$. As expounded in [28,43], the highest antighost number $I$ term should be strictly satisfied by $\gamma a_{1}^{(I)}=0$ and adopting the results developed in [28], since the Cauchy order of our system equals to two, we claim that $H_{I}(\delta|d)$ vanishes for $I>2$. Hence the expansion of (2.28) truncates at the antighost number two and we acquire
\begin{equation}
\begin{aligned}
a_{1}=a_{1}^{(0)} +a_{1}^{(1)} +a_{1}^{(2)}
\end{aligned}
\end{equation}
with $\gamma a_{1}^{(2)}=0$. Through a simple observation from (2.11) and (2.12), we assert that the Koszul-Tate differential $\delta$ lowers the antighost number whereas the exterior longitudinal derivative $\gamma$ keeps the antighost number. Inserting (2.29) into (2.27) and comparing the antighost number order by order, we immediately arrive at recursions for $a_{1}^{(i)}$
\begin{equation}
\begin{aligned}
\gamma a_{1}^{(2)}=0,\quad \delta a_{1}^{(2)}+\gamma a_{1}^{(1)}=\partial_{\mu}V^{\mu(1)}_{1},\quad \delta a_{1}^{(1)}+\gamma a_{1}^{(0)}=\partial_{\mu}V^{\mu(0)}_{1}
\end{aligned}
\end{equation}
which enable us to construct the explicit expression of $a_{1}$ and firstly, we find that the general form of the solution $a_{1}^{(2)}$ is given by
\begin{equation}
\begin{aligned}
a_{1}^{(2)}=\frac{1}{2}f^{a}_{bc}\eta^{*}_{a}\eta^{b}\eta^{c}
\end{aligned}
\end{equation}
here the constant $f^{a}_{bc}$ is antisymmetric with respect to the subscript
 $b$ and $c$, that is
\begin{equation}
\begin{aligned}
f^{a}_{bc}=-f^{a}_{cb}
 \end{aligned}
\end{equation}
 the precise meaning of these coefficients is explained below. Proceeding recursively, under this assumption by using the definition of $\gamma$ above, there is no difficulty in inferring
\begin{equation}
\begin{aligned}
\delta a_{1}^{(2)}=\frac{1}{2}\partial_{\mu}(-f^{a}_{bc}A^{*\mu}_{a}\eta^{b}\eta^{c})+\gamma(f^{a}_{bc}A^{*\mu}_{a}\eta^{b}A^{c}_{\mu}+f^{a}_{bc}B^{*\mu}_{a}\eta^{b}B^{c}_{\mu}+\phi_{i}^{*}\Gamma^{i}_{a}\eta^{a}+Z_{i}^{*}\bar{\Gamma}^{i}_{a}\eta^{a})
\end{aligned}
\end{equation}
it is necessary to remark here that an extra term $\phi_{i}^{*}\Gamma^{i}_{a}\eta^{a}+Z_{i}^{*}\bar{\Gamma}^{i}_{a}\eta^{a}$ come up in the formula (2.33), the conceptual reason is that although this term has no influence on the value of $\delta a_{1}^{(2)}$ but it can produce great significance on the derivation of the desired consistent deformation terms among the gauge and matter fields which will be discussed quickly. For the moment, by companion of (2.30) with (2.33), it is easy for us to get
\begin{equation}
\begin{aligned}
a_{1}^{(1)}=-f^{a}_{bc}A^{*\mu}_{a}\eta^{b}A^{c}_{\mu}-f^{a}_{bc}B^{*\mu}_{a}\eta^{b}B^{c}_{\mu}-\phi_{i}^{*}\Gamma^{i}_{a}\eta^{a}-Z_{i}^{*}\bar{\Gamma}^{i}_{a}\eta^{a}
\end{aligned}
\end{equation}
applying the above recursion successively, we obtain
\begin{equation}
\begin{aligned}
\delta a_{1}^{(1)}=&-f^{a}_{bc}(\partial_{\nu}F^{\mu\nu}_{a}+m^{2}\square B^{\mu}_{a}-m^{2}\partial^{\mu}\partial_{\nu}B^{\nu}_{a})\eta^{b}A^{c}_{\mu}-m^{2}f^{a}_{bc}(B^{\mu}_{a}+\partial_{\nu}F^{\nu\mu}_{a})\eta^{b}B^{c}_{\mu}\\
&+\delta(-\phi_{i}^{*}\Gamma^{i}_{a}-Z_{i}^{*}\bar{\Gamma}^{i}_{a})\eta^{a}
\end{aligned}
\end{equation}

The next task is to deduce the exact formula of antighost number zero part $a_{1}^{(0)}$, at first glance, in view of the fact that the action of $\gamma$ on $F_{\mu\nu}^{a}$ is trivial, namely $\gamma F_{\mu\nu}^{a}=0$ we are able to perform
\begin{equation}
\begin{aligned}
&\gamma(\frac{1}{2}f^{a}_{bc}F^{\mu\nu}_{a}A^{b}_{\mu}A^{c}_{\nu}-m^{2}f^{a}_{bc}A_{\mu}^{b}B_{\nu}^{c}F^{\mu\nu}_{a}-m^{2}f^{a}_{bc}\partial^{\mu}B^{\nu}_{a}A_{\mu}^{b}A_{\nu}^{c})\\
=&f^{a}_{bc}F^{\mu\nu}_{a}\partial_{\mu}\eta^{b}A^{c}_{\nu}-m^{2}f^{a}_{bc}\partial_{\mu}\eta^{b}B_{\nu}^{c}F^{\mu\nu}_{a}-m^{2}f^{a}_{bc}\partial^{\mu}B^{\nu}_{a}(\partial_{\mu}\eta^{b}A_{\nu}^{c}+A_{\mu}^{b}\partial_{\nu}\eta^{c})\\
\simeq&-f^{a}_{bc}(\eta^{b}\partial_{\mu}F^{\mu\nu}_{a}A^{c}_{\nu}+\eta^{b}F^{\mu\nu}_{a}\partial_{\mu}A^{c}_{\nu}-m^{2}\eta^{b}\partial_{\mu}B_{\nu}^{c}F^{\mu\nu}_{a}-m^{2}\eta^{b}B_{\nu}^{c}\partial_{\mu}F^{\mu\nu}_{a}\\
&-m^{2}\square B^{\nu}_{a}\eta^{b}A_{\nu}^{c}-m^{2}\partial_{\nu}\partial^{\mu}B^{\nu}_{a}A_{\mu}^{b}\eta^{c}-m^{2}\partial^{\mu}B^{\nu}_{a}\eta^{b}F_{\mu\nu}^{c})\\
=&f^{a}_{bc}\eta^{b}\partial_{\nu}F^{\mu\nu}_{a}A^{c}_{\mu}-\frac{1}{4}(f^{a}_{bc}+f^{c}_{ba})F^{\mu\nu}_{a}\eta^{b}F^{c}_{\mu\nu}+m^{2}(f^{a}_{bc}+f^{c}_{ba})\eta^{b}\partial_{\mu}B_{\nu}^{c}F^{\mu\nu}_{a}\\
&+m^{2}f^{a}_{bc}(\eta^{b}B_{\nu}^{c}\partial_{\mu}F^{\mu\nu}_{a}+\square B^{\nu}_{a}\eta^{b}A_{\nu}^{c}+\partial_{\nu}\partial^{\mu}B^{\nu}_{a}A_{\mu}^{b}\eta^{c})
\end{aligned}
\end{equation}
here $\simeq$ denotes the equivalence up to total derivative terms since they make no contributions to the integral in (2.30). At this stage, in order to construct the consistent interactions in the original theory, with the help of (2.30) we emphasize that the sum of (2.35) and (2.36) must be a total derivative term and therefore we should drop out the redundant terms, these are
\begin{equation}
\begin{aligned}
f^{a}_{bc}B^{\mu}_{a}\eta^{b}B^{c}_{\mu}=0,\quad\quad  (f^{a}_{bc}+f^{c}_{ba})F^{\mu\nu}_{a}\eta^{b}F^{c}_{\mu\nu}=0,\quad \quad (f^{a}_{bc}+f^{c}_{ba})\eta^{b}\partial_{\mu}B_{\nu}^{c}F^{\mu\nu}_{a}=0
\end{aligned}
\end{equation}
these vanishing conditions immediately give rise to the following relations
\begin{equation}
\begin{aligned}
f^{a}_{bc}=-f^{c}_{ba}
\end{aligned}
\end{equation}
moreover, as we have pointed out that the last term in (2.35) is also natural to demand to be a total derivative term which occurs only when
\begin{equation}
\begin{aligned}
\delta(-\phi_{i}^{*}\Gamma^{i}_{a}-Z_{i}^{*}\bar{\Gamma}^{i}_{a})=\partial_{\mu}J^{\mu}_{a}
\end{aligned}
\end{equation}
it is now enough to determine the $a_{1}^{(0)}$ as follows
\begin{equation}
\begin{aligned}
a_{1}^{(0)}=&\frac{1}{2}f^{a}_{bc}F^{\mu\nu}_{a}A^{b}_{\mu}A^{c}_{\nu}-m^{2}f^{a}_{bc}A_{\mu}^{b}B_{\nu}^{c}F^{\mu\nu}_{a}-m^{2}f^{a}_{bc}\partial^{\mu}B^{\nu}_{a}A_{\mu}^{b}A_{\nu}^{c}+J^{\mu}_{a}A^{a}_{\mu}
\end{aligned}
\end{equation}
collecting these results together, we eventually have gathered the basic ingredients to express the first-order deformation in the form of
\begin{equation}
\begin{aligned}
S_{1}=\int d^{4}x(f^{a}_{bc}&(\frac{1}{2}F^{\mu\nu}_{a}A^{b}_{\mu}A^{c}_{\nu}-m^{2}A_{\mu}^{b}B_{\nu}^{c}F^{\mu\nu}_{a}-m^{2}\partial^{\mu}B^{\nu}_{a}A_{\mu}^{b}A_{\nu}^{c}-A^{*\mu}_{a}\eta^{b}A^{c}_{\mu}\\
&-B^{*\mu}_{a}\eta^{b}B^{c}_{\mu}+\frac{1}{2}\eta^{*}_{a}\eta^{b}\eta^{c})-\phi_{i}^{*}\Gamma^{i}_{a}\eta^{a}-Z_{i}^{*}\bar{\Gamma}^{i}_{a}\eta^{a}+J^{\mu}_{a}A^{a}_{\mu})
\end{aligned}
\end{equation}

It is interesting to ask whether we can provide a concrete representation for the matter currents $J^{\mu}_{a}$, to capture this aspect, it seems inevitable to consider the second-order deformation $S_{2}=\int d^{4}xa_{2}$. Taking into account of the deformation equation $(S_{1},S_{1})+2(S_{0},S_{2})=0$ or as described before, in the local functional form this equation is understood in the sense of
\begin{equation}
\begin{aligned}
s_{11}+2sa_{2}=\partial_{\mu}V^{\mu}_{2}
\end{aligned}
\end{equation}
here $(S_{i},S_{j})=\int d^{4}xs_{ij}$. By the general arguments above, one is tempted to divide $s_{11}$ in accordance with the antighost number into three parts $s_{11}=s_{11}^{(0)}+s_{11}^{(1)}+s_{11}^{(2)}$ and utilize the canonical relations
\begin{equation}
\begin{aligned}
(A_{\mu}^{a}(x),A_{b}^{*\nu}(y))&=(A_{b}^{*\nu}(y),A_{\mu}^{a}(x))=-\delta^{a}_{b}\delta^{\nu}_{\mu}\delta^{4}(x-y),\\
(B_{\mu}^{a}(x),B_{b}^{*\nu}(y))&=(B_{b}^{*\nu}(y),B_{\mu}^{a}(x))=-\delta^{a}_{b}\delta^{\nu}_{\mu}\delta^{4}(x-y),\\
(\phi^{i}(x),\phi_{j}^{*}(y))&=(\phi_{j}^{*}(y),\phi^{i}(x))=-\delta^{i}_{j}\delta^{4}(x-y),\\
(Z^{i}(x),Z_{j}^{*}(y))&=(Z_{j}^{*}(y),Z^{i}(x))=-\delta^{i}_{j}\delta^{4}(x-y),\\
(\eta^{a}(x),\eta_{b}^{*}(y))&=(\eta_{b}^{*}(y),\eta^{a}(x))=-\delta^{a}_{b}\delta^{4}(x-y)
\end{aligned}
\end{equation}
after some cumbersome simplifications, the explicit expressions of the antighost number one and two are shown as
\begin{equation}
\begin{aligned}
s_{11}^{(1)}=&-(f^{a}_{ib}f^{i}_{cd}+f^{a}_{ic}f^{i}_{db}+f^{a}_{id}f^{i}_{bc})(A^{*\mu}_{a}\eta^{b}\eta^{c}A^{d}_{\mu}+B^{*\mu}_{a}\eta^{b}\eta^{c}B^{d}_{\mu})\\
&+(f^{a}_{bc}\Gamma^{i}_{a}-\frac{\delta^{R}\Gamma^{i}_{c}}{\delta\phi^{j}}\Gamma_{b}^{j}+\frac{\delta^{R}\Gamma^{i}_{b}}{\delta\phi^{j}}\Gamma_{c}^{j})\phi_{i}^{*}\eta^{b}\eta^{c}+(f^{a}_{bc}\bar{\Gamma}^{i}_{a}-\frac{\delta^{R}\bar{\Gamma}^{i}_{c}}{\delta Z^{j}}\bar{\Gamma}_{b}^{j}+\frac{\delta^{R}\bar{\Gamma}^{i}_{b}}{\delta Z^{j}}\bar{\Gamma}_{c}^{j})Z_{i}^{*}\eta^{b}\eta^{c},\\
s_{11}^{(2)}=&-\frac{1}{6}(f^{a}_{ib}f^{i}_{cd}+f^{a}_{ic}f^{i}_{db}+f^{a}_{id}f^{i}_{bc})\eta^{*}_{a}\eta^{b}\eta^{c}\eta^{d}
\end{aligned}
\end{equation}
in the similar spirit, we further expand $a_{2}$ as
\begin{equation}
\begin{aligned}
a_{2}=a_{2}^{(0)}+a_{2}^{(1)}+a_{2}^{(2)}
\end{aligned}
\end{equation}
here the antighost number of $a_{2}^{(i)}$ is $i$. In more detail, inserting (2.45) into (2.42), we are thus led to the equalities
\begin{equation}
\begin{aligned}
&s_{11}^{(2)}+2\gamma a_{2}^{(2)}=\partial_{\mu}k^{\mu(2)}_{2},\\
& s_{11}^{(1)}+2\delta a_{2}^{(2)}+2\gamma a_{2}^{(1)}=\partial_{\mu}k^{\mu(1)}_{1},\\ &s_{11}^{(0)}+2\delta a_{2}^{(1)}+2\gamma a_{2}^{(0)}=\partial_{\mu}k^{\mu(0)}_{1}
\end{aligned}
\end{equation}
it follows from (2.46) that through the comparison of both sides of equations, the $s_{11}^{(2)}$ and $2\gamma a_{2}^{(2)}$ should be total derivative terms which indeed is impossible only if [43,45]
\begin{equation}
\begin{aligned}
s_{11}^{(2)}=0,\quad \quad \quad a_{2}^{(2)}=0
\end{aligned}
\end{equation}
inspecting (2.44) we see that this happens only if
\begin{equation}
\begin{aligned}
&f^{a}_{ib}f^{i}_{cd}+f^{a}_{ic}f^{i}_{db}+f^{a}_{id}f^{i}_{bc}=0
\end{aligned}
\end{equation}
then, continuing this process and we find the terms $s_{11}^{(1)},a_{2}^{(1)}$ behave in the same way as
\begin{equation}
\begin{aligned}
s_{11}^{(1)}=0, \quad \quad a_{2}^{(1)}=0
\end{aligned}
\end{equation}
for each value of $i$, the above condition will give rise to a system of functional relations for the unknown $\Gamma^{i}_{a},\bar{\Gamma}^{i}_{a}$
\begin{equation}
\begin{aligned} f^{a}_{bc}\Gamma^{i}_{a}-\frac{\delta^{R}\Gamma^{i}_{c}}{\delta\phi^{j}}\Gamma_{b}^{j}+\frac{\delta^{R}\Gamma^{i}_{b}}{\delta\phi^{j}}\Gamma_{c}^{j}=0,\quad \quad f^{a}_{bc}\bar{\Gamma}^{i}_{a}-\frac{\delta^{R}\bar{\Gamma}^{i}_{c}}{\delta Z^{j}}\bar{\Gamma}_{b}^{j}+\frac{\delta^{R}\bar{\Gamma}^{i}_{b}}{\delta Z^{j}}\bar{\Gamma}_{c}^{j}=0
\end{aligned}
\end{equation}
for brevity, we just suppose the functionals $\Gamma_{a}^{i},\bar{\Gamma}_{a}^{i}$ taking the linear form, namely
\begin{equation}
\begin{aligned}
\Gamma_{a}^{i}=\Gamma_{aj}^{i}\phi^{j},\quad \quad \bar{ \Gamma}_{a}^{i}=\bar{\Gamma}_{aj}^{i}Z^{j}
\end{aligned}
\end{equation}
plugging these expressions into (2.50) we get
\begin{equation}
\begin{aligned}
\Gamma_{bj}^{i}\Gamma_{ck}^{j}-\Gamma_{cj}^{i}\Gamma_{bk}^{j}=f^{a}_{bc}\Gamma^{i}_{ak},\quad \quad \bar{\Gamma}_{bj}^{i}\bar{\Gamma}_{ck}^{j}-\bar{\Gamma}_{cj}^{i}\bar{\Gamma}_{bk}^{j}=f^{a}_{bc}\bar{\Gamma}^{i}_{ak}
\end{aligned}
\end{equation}
in principle, the accurate deformation equation (2.42) now could be carried out easily when we employ the identities (2.48) and (2.50) which signifies that the $s_{11}^{(1)},s_{11}^{(2)},a_{2}^{(1)},a_{2}^{(2)}$ will not show up.

The constraints between $G_{ij}$ and $\Gamma^{i}_{aj},\bar{\Gamma}^{i}_{aj}$ can be analyzed by exploiting the conditions (2.39), or more concretely we have
\begin{equation}
\begin{aligned}
&\delta(-\phi_{i}^{*}\Gamma^{i}_{a}-Z_{i}^{*}\bar{\Gamma}^{i}_{a})=\delta(-\phi_{i}^{*}\Gamma^{i}_{aj}\phi^{j}-Z_{i}^{*}\bar{\Gamma}^{i}_{aj}Z^{j})\\
=&\partial_{\mu}J_{a}^{\mu}+\frac{1}{2}(G_{ij}\Gamma_{al}^{i}\partial^{\mu}\phi^{j}\partial_{\mu}Z^{l}+G_{ij}\bar{\Gamma}_{al}^{i}\partial^{\mu}Z^{j}\partial_{\mu}\phi^{l})+\frac{1}{4}\alpha G_{il}\phi^{l}\Gamma^{i}_{aj}\phi^{j}\\
&+\frac{1}{4}\alpha G_{il}Z^{l}\bar{\Gamma}^{i}_{aj}Z^{j}-\frac{1}{4}\alpha G_{il}Z^{l}\Gamma^{i}_{aj}\phi^{j}-\frac{1}{4}\alpha G_{il}\phi^{l}\bar{\Gamma}^{i}_{aj}Z^{j}
\end{aligned}
\end{equation}
here the matter currents $J_{a}^{\mu}$ take the form of
\begin{equation}
\begin{aligned}
&J_{a}^{\mu}=-\frac{1}{2}(G_{il}\partial^{\mu}\phi^{l}\Gamma_{aj}^{i}Z^{j}+G_{il}\partial^{\mu}Z^{l}\bar{\Gamma}_{aj}^{i}\phi^{j})
\end{aligned}
\end{equation}
if we want to obtain a non-trivial local cohomology for Koszul-Tate differential $\delta$ at antighost number zero we should require that the right-hand side of (2.53) be a total derivative term which can be achieved only if
\begin{equation}
\begin{aligned}
G_{ij}\Gamma_{al}^{i}+G_{il}\Gamma_{aj}^{i}=0,\quad \quad G_{ij}\bar{\Gamma}_{al}^{i}+G_{il}\bar{\Gamma}_{aj}^{i}=0,\quad \quad G_{ij}\Gamma_{al}^{i}+G_{il}\bar{\Gamma}_{aj}^{i}=0
\end{aligned}
\end{equation}
under these constraints, the coefficients $\Gamma_{aj}^{i}$ and $\bar{\Gamma}_{aj}^{i}$ are no longer independent but in fact related by
\begin{equation}
\begin{aligned}
\Gamma_{aj}^{i}=\bar{\Gamma}_{aj}^{i}
\end{aligned}
\end{equation}

The Jacobi identity (2.48) shows strong evidence that it makes possible to rewrite the first-order deformation in the matrix form which turns out to be much convenient to explore the higher-order deformations. To see this, one introduces a set of matrix generators $(\Gamma_{a})_{ij}=\Gamma_{aj}^{i}$ satisfying the following commutation relations
\begin{equation}
\begin{aligned}
\left[\Gamma_{b},\Gamma_{c}\right]=f^{a}_{bc}\Gamma_{a}
\end{aligned}
\end{equation}
essentially by imposing the normalization condition
\begin{equation}
\begin{aligned}
\mathrm{tr}(\Gamma_{a}\Gamma_{b})=\delta_{ab}
\end{aligned}
\end{equation}
with the suitable choices of $f^{a}_{bc}$ which will further lead to drastic simplifications of our derivations and the Jacobi identity
\begin{equation}
\begin{aligned}
\left[\Gamma_{a},\left[\Gamma_{b},\Gamma_{c}\right]\right]+\left[\Gamma_{b},\left[\Gamma_{c},\Gamma_{a}\right]\right]+\left[\Gamma_{c},\left[\Gamma_{a},\Gamma_{b}\right]\right]=0
\end{aligned}
\end{equation}
follows as a consequence of (2.48). We are aware of the fact that these $\Gamma_{a}$ constitute the basis of generators of the representation of some Lie algebra $\mathfrak{g}$ and the $f^{a}_{bc}$ can be interpreted as the structure constant coefficients. Also the relations (2.48) and (2.52) simply revel that the deformed gauge transformations of the gauge and matter fields generate the Lie algebras with the same structure constants. From now on, equipped with this information it is reasonable to assume that the fields in Lagrangian take values in the form of Lie algebra $\mathfrak{g}$, that is we would like to make the replacements
\begin{equation}
\begin{aligned}
A^{a}_{\mu}\rightarrow A_{\mu}=A^{a}_{\mu}\Gamma_{a},\quad B^{a}_{\mu}\rightarrow B_{\mu}=B^{a}_{\mu}\Gamma_{a},\quad F^{a}_{\mu\nu}\rightarrow F^{\mu\nu}=F^{a}_{\mu\nu}\Gamma_{a},\quad \eta^{a}\rightarrow \eta=\eta^{a}\Gamma_{a}
\end{aligned}
\end{equation}
at the same time we adopt the notation $\phi,Z$ for the column vectors of the scalar fields
\begin{equation}
\begin{aligned}
\phi=(\phi^{1},\phi^{2},......,\phi^{M})^{T},\quad\quad  Z=(Z^{1},Z^{2},......,Z^{M})^{T}
 \end{aligned}
\end{equation}
 and of course the row vectors $\phi^{T},Z^{T}$ are the transposes of $\phi,Z$. The relations (2.55) are then expressible in terms of the matrices $\Gamma_{a}$ as
\begin{equation}
\begin{aligned}
G\Gamma_{a}=-(G\Gamma_{a})^{T}=-\Gamma_{a}^{T}G,\quad \quad \Gamma_{a}=\bar{\Gamma}_{a}
\end{aligned}
\end{equation}
here $G^{T}=G$ since we suppose the matrix $G$ is symmetric. Upon combining $A^{a}_{\mu}$ with $\Gamma_{a}$ we learn that
\begin{equation}
\begin{aligned}
GA_{\mu}=-(GA_{\mu})^{T}=-A_{\mu}^{T}G,\quad \quad G\eta=-(G\eta)^{T}=-\eta^{T}G
\end{aligned}
\end{equation}
which will be useful in the calculations below. Employing (2.54), it is not difficult to find that the matter currents $J^{\mu}_{a}$ coupled to gauge fields $A^{a}_{\mu}$ can be effectively rewritten as
\begin{equation}
\begin{aligned}
J^{\mu}_{a}A^{a}_{\mu}=-\frac{1}{2}((\partial^{\mu}\phi)^{T}GA_{\mu}Z+(\partial^{\mu}Z)^{T}GA_{\mu}\phi)
\end{aligned}
\end{equation}
and in fact, the conserved currents $J^{\mu}_{a}$ (2.54) could be received from the global invariance
\begin{equation}
\begin{aligned}
\Delta\phi=\Gamma_{a}\phi+\xi,\quad \quad \quad \Delta Z=\Gamma_{a}Z+\xi
\end{aligned}
\end{equation}
of the real scalar fields in Lagrangian (2.4) according to the Noether's theorem, here $\xi$ is arbitrary constant column vector.

As a result, the above $\mathfrak{g}$-valued fields allow us to write down the first-order deformation $S_{1}$ in a more compact and familiar form
\begin{equation}
\begin{aligned}
S_{1}=\mathrm{tr}\int d^{4}x&(\frac{1}{2}F^{\mu\nu}\left[A_{\mu},A_{\nu}\right]-m^{2}\left[A_{\mu},B_{\nu}\right]F^{\mu\nu}-m^{2}\partial_{\mu}B_{\nu}\left[A^{\mu},A^{\nu}\right]\\
&-A^{*\mu}\left[\eta,A_{\mu}\right]-\frac{1}{2}(\partial^{\mu}\phi)^{T}GA_{\mu}Z-\frac{1}{2}(\partial^{\mu}Z)^{T}GA_{\mu}\phi\\
&-B^{*\mu}\left[\eta,B_{\mu}\right]-\phi^{*T}\eta\phi-Z^{*T}\eta Z+\frac{1}{2}\eta^{*}\left[\eta,\eta\right])
 \end{aligned}
\end{equation}
the procedure of construction of the first-order deformation described above is most general and seems to be very meaningful, that requires further study to seek how it works in the determination of the total deformed master action. In the next section we will mainly calculate the higher-order deformations within this formalism and illustrate its applications in other specific situations, including the complex scalar fields and Dirac spinor fields.

\subsection{Second-order deformation}
 In this section we concentrate on the computation of the seconde-order deformation  $S_{2}$ and to begin with, for later convenience, we decompose the $S_{1}$ into three parts
\begin{equation}
\begin{aligned}
S_{1}^{g}=&\mathrm{tr}\int d^{4}x(\frac{1}{2}F^{\mu\nu}\left[A_{\mu},A_{\nu}\right]-m^{2}\left[A_{\mu},B_{\nu}\right]F^{\mu\nu}-m^{2}\partial_{\mu}B_{\nu}\left[A^{\mu},A^{\nu}\right]),\\
S_{1}^{m}=&\mathrm{tr}\int d^{4}x\frac{1}{2}(-(\partial^{\mu}\phi)^{T}GA_{\mu}Z-(\partial^{\mu}Z)^{T}GA_{\mu}\phi),\\
S_{1}^{A}=&\mathrm{tr}\int d^{4}x(-A^{*\mu}\left[\eta,A_{\mu}\right]-B^{*\mu}\left[\eta,B_{\mu}\right]-\phi^{*T}\eta\phi-Z^{*T}\eta Z+\frac{1}{2}\eta^{*}\left[\eta,\eta\right])
 \end{aligned}
\end{equation}
here $S_{1}^{g}$ is the first-order deformation corresponds to the pure gauge part, $S_{1}^{m}$ is the first-order deformation of the matter fields part which can be regarded as the couplings among the gauge and matter fields and the $S_{1}^{A}$ only contains the antifields terms. For the purpose of obtaining the second-order deformations, we realize that the evaluation of $(S_{1},S_{1})$ only originates from the antibracket between $(S_{1}^{g},S_{1}^{A})$ as well as $(S_{1}^{m},S_{1}^{A})$ since $(S_{1}^{g},S_{1}^{m})$ vanishes automatically. Therefore in practice, this simple observation forces us to divide the second-order deformation into $S_{2}=S_{2}^{g}+S_{2}^{m}$ and thus it is suffice to tackle the equations
\begin{equation}
\begin{aligned}
 (S_{1}^{g},S_{1}^{A})+sS_{2}^{g}=0,\quad \quad \quad  (S_{1}^{m},S_{1}^{A})+sS_{2}^{m}=0
 \end{aligned}
\end{equation}
for the sake of concreteness, we intend to compute $(S_{1}^{g},S_{1}^{A})$ and paying attention that for arbitrary function $f(x)$, we have the following fact of the Dirac's function
\begin{equation}
\begin{aligned}
\int_{-\infty}^{\infty}f(x)\frac{d^{n}}{d^{n}x}\delta(x-c)=(-1)^{n}\left[\frac{d^{n}}{d^{n}x}f(x)\right]_{x=c}
\end{aligned}
\end{equation}
then it follows that
\begin{equation}
\begin{aligned}
&(\mathrm{tr}\int d^{4}x\frac{1}{2}F^{\mu\nu}\left[A_{\mu},A_{\nu}\right],-\mathrm{tr}\int d^{4}yA^{*j}\left[\eta,A_{j}\right])\\
=&\frac{1}{2}\mathrm{tr}\int d^{4}x\eta(\left[A_{\mu},\left[A_{\nu},F^{\mu\nu}\right]\right]+\left[A_{\nu},\left[F^{\mu\nu},A_{\mu}\right]\right]-\left[A^{\nu},\partial^{\mu}\left[A_{\mu},A_{\nu}\right]\right]+\left[A^{\mu},\partial^{\nu}\left[A_{\mu},A_{\nu}\right]\right])\\
=&\frac{1}{2}\mathrm{tr}\int d^{4}x\eta(-\left[F_{\mu\nu},\left[A^{\mu},A^{\nu}\right]\right]-\left[A^{\nu},\partial^{\mu}\left[A_{\mu},A_{\nu}\right]\right]+\left[A^{\mu},\partial^{\nu}\left[A_{\mu},A_{\nu}\right]\right])
\end{aligned}
\end{equation}
in the above calculation, use has been made of the following property of trace
\begin{equation}
\begin{aligned}
\mathrm{tr}(\left[A,B\right]C)=\mathrm{tr}(A\left[B,C\right])
\end{aligned}
\end{equation}
together with the help of the Jacobi identity. For the remaining terms in (2.68), applying the property of the Dirac's function again we have
\begin{equation}
\begin{aligned}
&(\mathrm{tr}\int d^{4}x(-\left[A_{\mu},B_{\nu}\right]F^{\mu\nu}-\partial_{\mu}B_{\nu}\left[A^{\mu},A^{\nu}\right]),-\mathrm{tr}\int d^{4}y(A^{*\mu}\left[\eta,A_{\mu}\right]+B^{*\mu}\left[\eta,B_{\mu}\right]))\\
=&\mathrm{tr}\int d^{4}x\eta(\left[A^{\nu},\partial^{\mu}\left[A_{\mu},B_{\nu}\right]\right]-\left[A_{\mu},\left[B_{\nu},F^{\mu\nu}\right]\right]-\left[A^{\mu},\partial^{\nu}\left[A_{\mu},B_{\nu}\right]\right]-\left[B_{\nu},\left[F^{\mu\nu},A_{\mu}\right]\right]\\
&+\left[B_{\nu},\partial_{\mu}\left[A^{\mu},A^{\nu}\right]\right]-\left[A^{\mu},\left[A^{\nu},\partial_{\mu}B_{\nu}\right]\right]-\left[A^{\nu},\left[\partial_{\mu}B_{\nu},A^{\mu}\right]\right])
\end{aligned}
\end{equation}
on the other hand, taking the partial integration we notice that
\begin{equation}
\begin{aligned}
&s(\mathrm{tr}\int d^{4}x-\frac{1}{4}\left[A_{\mu},A_{\nu}\right]\left[A^{\mu},A^{\nu}\right])\\
=&-\frac{1}{2}\mathrm{tr}\int d^{4}x(\left[\partial_{\mu}\eta,A_{\nu}\right]+\left[A_{\mu},\partial_{\nu}\eta\right])\left[A^{\mu},A^{\nu}\right]\\
=&\frac{1}{2}\mathrm{tr}\int d^{4}x\eta(\left[F_{\mu\nu},\left[A^{\mu},A^{\nu}\right]\right]+\left[A_{\nu},\partial_{\mu}\left[A^{\mu},A^{\nu}\right]\right]+\left[\partial_{\nu}\left[A^{\mu},A^{\nu}\right],A_{\mu}\right])
\end{aligned}
\end{equation}
in the similar way we acquire
\begin{equation}
\begin{aligned}
&s(\mathrm{tr}\int d^{4}x\left[A_{\mu},B_{\nu}\right]\left[A^{\mu},A^{\nu}\right])\\
=&-\mathrm{tr}\int d^{4}x\eta(\left[\partial_{\mu}B_{\nu},\left[A^{\mu},A^{\nu}\right]\right]+\left[B_{\nu},\partial^{\mu}\left[A^{\mu},A^{\nu}\right]\right]+\left[F^{\mu\nu},\left[A_{\mu},B_{\nu}\right]\right]\\
&+\left[A^{\nu},\partial^{\mu}\left[A_{\mu},B_{\nu}\right]\right]+\left[\partial^{\nu}\left[A_{\mu},B_{\nu}\right],A^{\mu}\right])
\end{aligned}
\end{equation}
from these explicit expressions and by using the Jacobi identity
\begin{equation}
\begin{aligned}
\left[A_{\mu},\left[B_{\nu},F^{\mu\nu}\right]\right]+\left[B_{\nu},\left[F^{\mu\nu},A_{\mu}\right]\right]+\left[F^{\mu\nu},\left[A_{\mu},B_{\nu}\right]\right]=0
\end{aligned}
\end{equation}
as well as
\begin{equation}
\begin{aligned}
\left[A^{\mu},\left[A^{\nu},\partial_{\mu}B_{\nu}\right]\right]+\left[A^{\nu},\left[\partial_{\mu}B_{\nu},A^{\mu}\right]\right]+\left[\partial_{\mu}B_{\nu},\left[A^{\mu},A^{\nu}\right]\right]=0
\end{aligned}
\end{equation}
we find that the sum of above $s$-exact terms is in agreement with $-(S_{1}^{g},S_{1}^{A})$ and consequently the solution of deformation equations (2.68) is given by
\begin{equation}
\begin{aligned}
S_{2}^{g}=&\mathrm{tr}\int d^{4}x(-\frac{1}{4}\left[A_{\mu},A_{\nu}\right]\left[A^{\mu},A^{\nu}\right]+m^{2}\left[A_{\mu},B_{\nu}\right]\left[A^{\mu},A^{\nu}\right])
\end{aligned}
\end{equation}

We now turn the attention to the investigation of $(S_{1}^{m},S_{1}^{A})$ which is determined from the canonical relations between the pairs $(A_{\mu},\phi,Z)$ and $(A^{*\mu},\phi^{*},Z^{*})$, and actually we have
\begin{equation}
\begin{aligned}
&(\int d^{4}xJ_{a}^{\mu}A_{\mu}^{a},-\mathrm{tr}\int d^{4}yA^{*\mu}\left[\eta,A_{\mu}\right])\\
=&-\frac{1}{2}\int d^{4}x((\partial^{\mu}\phi)^{T}G\left[\eta,A_{\mu}\right]Z+(\partial^{\mu}Z)^{T}G\left[\eta,A_{\mu}\right]\phi)
\end{aligned}
\end{equation}
along with
\begin{equation}
\begin{aligned}
&(\int d^{4}xJ_{a}^{\mu}A_{\mu}^{a},-\int d^{4}y(\phi^{*T}\eta\phi+Z^{*T}\eta Z))\\
=&\frac{1}{2}\int d^{4}x\left[(\eta\phi)^{T}G\partial^{\mu}(A_{\mu}Z)-(\partial^{\mu}\phi)^{T}GA_{\mu}\eta Z+(\eta Z)^{T}G\partial^{\mu}(A_{\mu}\phi)-(\partial^{\mu}Z)^{T}GA_{\mu}\eta \phi\right]
\end{aligned}
\end{equation}
which thus yield the following result
\begin{equation}
\begin{aligned}
(S_{1}^{m}+S_{1}^{A},S_{1}^{m}+S_{1}^{A})=&\int d^{4}x((\eta\phi)^{T}G\partial^{\mu}(A_{\mu}Z)-(\partial^{\mu}\phi)^{T}G\eta A_{\mu}Z\\
&+(\eta Z)^{T}G\partial^{\mu}(A_{\mu}\phi)-(\partial^{\mu}Z)^{T}G\eta A_{\mu}\phi)
\end{aligned}
\end{equation}
on the other hand, after a straightforward computation we obtain
\begin{equation}
\begin{aligned}
&s(\int d^{4}x(A_{\mu}\phi)^{T}GA^{\mu}Z)\\
=&\int d^{4}x((\partial_{\mu}\eta\phi)^{T}GA^{\mu}Z+(A_{\mu}\phi)^{T}G\partial^{\mu}\eta Z)\\
=&-\int d^{4}x((\eta\partial^{\mu}\phi)^{T}G A_{\mu}Z+(\eta\phi)^{T}G\partial^{\mu}(A_{\mu}Z)+\partial^{\mu}(A_{\mu} \phi)^{T}G\eta Z+(A_{\mu}\phi)^{T}G\eta \partial^{\mu}Z)
\end{aligned}
\end{equation}
by making using of the fact for $\eta^{T}$ as shown in (2.63), one asserts that
\begin{equation}
\begin{aligned}
(\eta\partial^{\mu}\phi)^{T}G=(\partial^{\mu}\phi)^{T}\eta^{T}G=-(\partial^{\mu}\phi)^{T}G\eta
\end{aligned}
\end{equation}
similarly for $(\eta\partial^{\mu}Z)^{T}G$ and comparing (2.80) with (2.81), we conclude that the solution of (2.68) for couplings part can be cast in the form
\begin{equation}
\begin{aligned}
S_{2}^{m}=&\frac{1}{2}\int d^{4}x(A_{\mu}\phi)^{T}GA^{\mu}Z
\end{aligned}
\end{equation}
finally putting (2.77) and (2.83) together, we get the second-order deformation $S_{2}=S_{2}^{g}+S_{2}^{m}$ which presents as
\begin{equation}
\begin{aligned}
S_{2}=&\mathrm{tr}\int d^{4}x(-\frac{1}{4}\left[A_{\mu},A_{\nu}\right]\left[A^{\mu},A^{\nu}\right]+m^{2}\left[A_{\mu},B_{\nu}\right]\left[A^{\mu},A^{\nu}\right]+\frac{1}{2}(A_{\mu}\phi)^{T}GA^{\mu}Z)
\end{aligned}
\end{equation}

\subsection{Higher-order deformation}
We now proceed to the evaluation of the third-order deformation at $g^{3}$ and as we have already said, in order to solve the deformation equation $(S_{1},S_{2})+sS_{3}=0$ we wish to compute the $(S_{1},S_{2})$. Firstly, it is interesting to recognize that by means of the Jacobi identity we simply infer
\begin{equation}
\begin{aligned}
&(-\mathrm{tr}\int d^{4}xA^{*\mu}\left[\eta,A_{\mu}\right],\mathrm{tr}\int d^{4}y(-\frac{1}{4}\left[A_{\mu},A_{\nu}\right]\left[A^{\mu},A^{\nu}\right]))=0\\
\end{aligned}
\end{equation}
analogously a direct calculation shows that
\begin{equation}
\begin{aligned}
&(-\mathrm{tr}\int d^{4}x(A^{*\nu}\left[\eta,A_{\nu}\right]+B^{*\mu}\left[\eta,B_{\mu}\right]),\mathrm{tr}\int d^{4}y\left[A_{\mu},B_{\nu}\right]\left[A^{\mu},A^{\nu}\right])\\
=&\mathrm{tr}\int d^{4}x\eta(\left[A_{\mu},\left[B_{\nu},\left[A^{\mu},A^{\nu}\right]\right]\right]+\left[A^{\mu},\left[A^{\nu},\left[A_{\mu},B_{\nu}\right]\right]\right]+\left[A^{\nu},\left[\left[A_{\mu},B_{\nu}\right],A^{\mu}\right]\right]\\
&+\left[B_{\nu},\left[\left[A^{\mu},A^{\nu}\right],A_{\mu}\right]\right])\\
=&\mathrm{tr}\int d^{4}x\eta(\left[A^{\mu},\left[A^{\nu},\left[A_{\mu},B_{\nu}\right]\right]\right]+\left[A^{\nu},\left[\left[A_{\mu},B_{\nu}\right],A^{\mu}\right]\right]-\left[\left[A^{\mu},A^{\nu}\right],\left[A_{\mu},B_{\nu}\right]\right])\\
=&0
\end{aligned}
\end{equation}
these consequences imply that
\begin{equation}
\begin{aligned}
(S_{1},S_{2}^{g})=0
\end{aligned}
\end{equation}
in addition, according to our discussion of the second-order deformation about the couplings part we need to calculate
\begin{equation}
\begin{aligned}
(-\int d^{4}y(\phi^{*T}\eta\phi+Z^{*T}\eta Z),S_{2}^{m})=&\frac{1}{2}\int d^{4}x(\phi^{T}G\eta A_{\mu}A^{\mu}Z-\phi^{T}GA_{\mu}A^{\mu}\eta Z)
\end{aligned}
\end{equation}
here we use property (2.63) again. Similarly, without efforts we deduce
\begin{equation}
\begin{aligned}
(-\mathrm{tr}\int d^{4}xA^{*\mu}\left[\eta,A_{\mu}\right],S_{2}^{m})=&\frac{1}{2}\int d^{4}x(\phi^{T}G A_{\mu}A^{\mu}\eta Z-\phi^{T}G\eta A_{\mu}A^{\mu} Z)
\end{aligned}
\end{equation}
certainly one can verify that
\begin{equation}
\begin{aligned}
(S_{1},S_{2}^{m})=0
\end{aligned}
\end{equation}
after substituting these results into the third-order deformation equation (2.21), we immediately assert that $S_{3}=0$ and furthermore, the other higher-order deformations can be fixed by the suitable choices of $S_{i}=0$ for $i\geq4$. At this point, we obtain the total solution of the deformation master equations of the reduced Lagrangian density that admits the form of
\begin{equation}
\begin{aligned}
S=&\mathrm{tr}\int d^{4}x(-\frac{1}{4}F_{\mu\nu}F^{\mu\nu}-\frac{m^{2}}{2}B_{\mu}B^{\mu}+m^{2}\partial_{\mu}B_{\nu}F^{\mu\nu}+\frac{1}{2}(\partial_{\mu}\phi)^{T}G\partial^{\mu}Z+\frac{1}{8}\alpha\phi^{T}G\phi\\
&+\frac{1}{8}\alpha Z^{T}GZ-\frac{1}{4}\alpha\phi^{T}GZ+A^{*\mu}\partial_{\mu}\eta+g(\frac{1}{2}F^{\mu\nu}\left[A_{\mu},A_{\nu}\right]-m^{2}\left[A_{\mu},B_{\nu}\right]F^{\mu\nu}\\
&-m^{2}\partial_{\mu}B_{\nu}\left[A^{\mu},A^{\nu}\right]-\frac{1}{2}(\partial^{\mu}\phi)^{T}GA_{\mu}Z-\frac{1}{2}(\partial^{\mu}Z)^{T}GA_{\mu}\phi-A^{*\mu}\left[\eta,A_{\mu}\right]\\
&-B^{*\mu}\left[\eta,B_{\mu}\right]+\frac{1}{2}\eta^{*}\left[\eta,\eta\right])+g^{2}(-\frac{1}{4}\left[A_{\mu},A_{\nu}\right]\left[A^{\mu},A^{\nu}\right]+m^{2}\left[A_{\mu},B_{\nu}\right]\left[A^{\mu},A^{\nu}\right]\\
&+\frac{1}{2}(A_{\mu}\phi)^{T}GA^{\mu}Z))
\end{aligned}
\end{equation}

To simplify this expression, let us introduce the field strength $\mathcal{F}_{\mu\nu}$, covariant derivative $D_{\mu}$ and the operator $d_{\mu}$
\begin{equation}
\begin{aligned}
\mathcal{F}_{\mu\nu}=F_{\mu\nu}-g\left[A_{\mu},A_{\nu}\right],\quad \quad D_{\mu}=\partial_{\mu}-g\left[A_{\mu},\quad \right],\quad\quad d_{\mu}=\partial_{\mu}-gA_{\mu}
\end{aligned}
\end{equation}
in this way, the antighost number zero part $\bar{S}_{0}\left[A_{\mu},\phi,Z\right]$ in the above deformed master action can be expressed as follows
\begin{equation}
\begin{aligned}
\bar{S}_{0}\left[A_{\mu},\phi,Z\right]=&\mathrm{tr}\int d^{4}x(-\frac{1}{4}\mathcal{F}_{\mu\nu}\mathcal{F}^{\mu\nu}-\frac{m^{2}}{2}B_{\mu}B^{\mu}+m^{2}D_{\mu}B_{\nu}\mathcal{F}^{\mu\nu}\\
&+\frac{1}{2}(d_{\mu}\phi)^{T}Gd^{\mu}Z+\frac{1}{8}\alpha\phi^{T}G\phi+\frac{1}{8}\alpha Z^{T}GZ-\frac{1}{4}\alpha\phi^{T}GZ)
\end{aligned}
\end{equation}
with the help of the equations of motion of auxiliary fields
\begin{equation}
\begin{aligned}
B^{\nu}=-D_{\mu}\mathcal{F}^{\mu\nu},\quad \quad  \quad Z=\phi+\frac{2}{\alpha}d_{\mu}d^{\mu}\phi
\end{aligned}
\end{equation}
 we are capable of converting this Lagrangian into an equivalent form
\begin{equation}
\begin{aligned}
\bar{S}_{0}=\mathrm{tr}\int d^{4}x(&-\frac{1}{4}\mathcal{F}_{\mu\nu}\mathcal{F}^{\mu\nu}+\frac{m^{2}}{2}D_{\mu}\mathcal{F}^{\mu\nu}D^{\lambda}\mathcal{F}_{\lambda\nu}+\frac{1}{2}(d_{\mu}\phi)^{T}Gd^{\mu}\phi\\
&-\frac{1}{2\alpha}(d_{\mu}d^{\mu}\phi)^{T}Gd^{\mu}d_{\mu}\phi)\\
\end{aligned}
\end{equation}
now comparing (2.1) with (2.95), it is evident to see that the corresponding couplings are induced through the replacement of the Abelian curvatures by non-Abelian ones and the usual ordinary derivatives by the covariant ones, together with some possible couplings terms among gauge and scalar fields which are necessary for consistency. This can also be seen by expanding the expressions of the power series of parameter $g$. In conclusion, the action (2.95) provides all the necessary information on the Lagrangian density of the interacting theory among the gauge and matter fields which can be regarded as the non-Abelian extension of the free higher derivative Yang-Mills gauge theory with matter fields, while the original local symmetries turn out to be the following non-Abelian gauge transformations
\begin{equation}
\begin{aligned}
\delta_{\xi}A_{\mu}^{a}=\partial_{\mu}\xi^{a}-gf^{a}_{bc}A_{\mu}^{b}\xi^{c},\quad \quad \quad \delta_{\xi}\phi^{i}=-g\Gamma^{i}_{aj}\phi^{j}\xi^{a}
\end{aligned}
\end{equation}
for arbitrary functions $\xi^{a}$ and the parameter $g$ is interpreted as the couplings constant among the gauge and matter fields. As explained in [54], when the free dynamics (2.1) is stable, the energy of system (2.95) including self-interacting and coupling terms can still have a local minimum in a neighborhood of zero solution and hence such theories are also considered as physically acceptable models which could be studied by means of perturbation expansion.

Consequently, the total solutions of the deformed master action can be formulate as (including the non-minimal term)
\begin{equation}
\begin{aligned}
S=&\mathrm{tr}\int d^{4}x(\bar{S}_{0}+A^{*\mu}\partial_{\mu}\eta-gA^{*\mu}\left[\eta,A_{\mu}\right]-gB^{*\mu}\left[\eta,B_{\mu}\right]+\frac{1}{2}g\eta^{*}\left[\eta,\eta\right]+\bar{\eta}^{*}\lambda)
\end{aligned}
\end{equation}
in order to evaluate the path integral of this non-Abelian coupling system, let us choose the following gauge-fixing fermion
\begin{equation}
\begin{aligned}
\Psi=\mathrm{tr}\int d^{4}x (\bar{\eta}n^{\mu}A_{\mu})
\end{aligned}
\end{equation}
here $n_{\mu}$ is an arbitrary constant vector and from (2.16), the non-trivial values of the antighost fields are easy to calculated as
\begin{equation}
\begin{aligned}
A_{\mu}^{*}=-\bar{\eta}n_{\mu},\quad \quad\bar{\eta}^{*}=-n^{\mu}A_{\mu}
\end{aligned}
\end{equation}
then path integral turns out to be
 \begin{equation}
\begin{aligned}
Z=\int\left[DA_{\mu}\right]\left[DB_{\mu}\right]\left[D\phi\right]\left[DZ\right]\left[D\eta\right]\left[D\bar{\eta}\right]\left[D\lambda\right]\mathrm{exp}\frac{i}{\hbar}(\bar{S}_{0}+\int d^{4}x(-\bar{\eta}n^{\mu}\bar{D}_{\mu}\eta-n^{\mu}A_{\mu}\lambda))
\end{aligned}
\end{equation}
here $\bar{D}_{\mu}=\partial_{\mu}+g\left[A_{\mu},\quad \right]$ and it is obvious to see that integral over the auxiliary field $\lambda$ will give us the usual axial gauge $n^{\mu}A_{\mu}=0$.
\section{Massive complex scalar fields}
In this section, we consider the following free Lagrangian density between the Abelian gauge fields and a set of complex massive scalar fields $(\varphi^{i},\bar{\varphi}_{i})$ with higher-order derivative terms
\begin{equation}
\begin{aligned}
\mathcal{L}=-\frac{1}{4}F_{\mu\nu}^{a}F^{\mu\nu}_{a}+\frac{m^{2}}{2}\partial_{\mu}F^{\mu\nu}_{a}\partial^{\lambda}F_{\lambda\nu}^{a}+\partial_{\mu}\varphi^{i}\partial^{\mu}\bar{\varphi}_{i}-\frac{1}{\alpha}\square\varphi^{i}\square\bar{\varphi}_{i}-M^{2}\varphi^{i}\bar{\varphi}_{i}
\end{aligned}
\end{equation}
here $M$ is some constant. The dynamic equations of motion for matter fields $(\varphi^{i},\bar{\varphi}_{i})$ associated to the Lagrangian density are
\begin{equation}
\begin{aligned}
\square(\square+\alpha)\varphi^{i}+\alpha M^{2}\varphi^{i}=0,\quad \quad \square(\square+\alpha)\bar{\varphi}_{i}+\alpha M^{2}\bar{\varphi}_{i}=0
\end{aligned}
\end{equation}
then by the standard approach of order reduction method, we introduce a set of complex auxiliary scalar fields $(Z^{i},\bar{Z}_{i})$ and the above Lagrangian can be equivalently recast in the form of
\begin{equation}
\begin{aligned}
\mathcal{\tilde{L}}=&-\frac{1}{4}F_{\mu\nu}^{a}F^{\mu\nu}_{a}-\frac{m^{2}}{2}B^{a}_{\mu}B^{\mu }_{a}+m^{2}\partial_{\mu}B_{\nu}^{a}F^{\mu\nu}_{a}+\frac{1}{2}(\partial_{\mu}\varphi^{i}\partial^{\mu}\bar{Z}_{i}+\partial^{\mu}\bar{\varphi}_{i}\partial_{\mu}Z^{i})\\
&+\frac{1}{4}\alpha\varphi^{i}\bar{\varphi}_{i}+\frac{1}{4}\alpha Z^{i}\bar{Z}_{i}-\frac{1}{4}\alpha(\varphi^{i} \bar{Z}_{i}+\bar{\varphi}_{i} Z^{i})-M^{2}\varphi^{i}\bar{\varphi}_{i}
\end{aligned}
\end{equation}
with the equations of motion
\begin{equation}
\begin{aligned}
&(1+2\frac{\square}{\alpha})\varphi^{i}=Z^{i},\quad \quad (1+2\frac{\square}{\alpha})\bar{\varphi}_{i}=\bar{Z}_{i},\\
&(2\square+\alpha)Z^{i}=\alpha\varphi^{i}-4M^{2}\varphi^{i},\quad \quad  (2\square+\alpha)\bar{Z}_{i}=\alpha\bar{\varphi}_{i}-4M^{2}\bar{\varphi}_{i}
\end{aligned}
\end{equation}
for complex scalar and auxiliary fields. As pointed out previously, the Lagrangian density (3.3) is invariant under the local gauge transformations
\begin{equation}
\begin{aligned}
\Delta_{\lambda} A_{\mu}^{a}=\partial_{\mu}\lambda^{a},\quad \Delta_{\lambda} \varphi^{i}=0,\quad \Delta_{\lambda} \bar{\varphi}_{i}=0,\quad \Delta_{\lambda}  Z^{i}=0,\quad \Delta_{\lambda} \bar{Z}_{i}=0
\end{aligned}
\end{equation}

It is convenient to list just the action of the Koszul-Tate differential on the generators $\varphi^{*}_{i},\bar{\varphi}^{*i},Z^{*}_{i},\bar{Z}^{*i}$ in the BRST complex due to the other actions are all trivial
\begin{equation}
\begin{aligned}
&\delta\varphi^{*}_{i}=\frac{1}{2}(\square \bar{Z}_{i}-\frac{1}{2}\alpha\bar{\varphi}_{i}+\frac{1}{2}\alpha \bar{Z}_{i})+M^{2}\bar{\varphi}_{i},\quad \delta Z^{*}_{i}=\frac{1}{2}(\square \bar{\varphi}_{i}-\frac{1}{2}\alpha \bar{Z}_{i}+\frac{1}{2}\alpha \bar{\varphi}_{i}),\\
&\delta\bar{\varphi}^{*i}=\frac{1}{2}(\square Z^{i}-\frac{1}{2}\alpha\varphi^{i}+\frac{1}{2}\alpha Z^{i})+M^{2}\varphi^{i},\quad \delta \bar{Z}^{*i}=\frac{1}{2}(\square \varphi^{i}-\frac{1}{2}\alpha Z^{i}+\frac{1}{2}\alpha\varphi^{i})
\end{aligned}
\end{equation}
proceeding as before, bearing in mind of (2.39) and now we need four sets of $i\Gamma^{i}_{aj},i\bar{\Gamma}_{ai}^{j},iT^{i}_{aj},i\bar{T}^{j}_{ai}$ corresponding to $\varphi^{j},\bar{\varphi}_{j},Z^{j},\bar{Z}_{j}$ respectively. From (3.6) we have
\begin{equation}
\begin{aligned}
&\delta(-i\varphi_{i}^{*}\Gamma^{i}_{aj}\varphi^{j}-i\bar{\varphi}^{*i}\bar{\Gamma}_{ai}^{j}\bar{\varphi}_{j}-iZ_{i}^{*}T^{i}_{aj}Z^{j}-i\bar{Z}^{*i}\bar{T}^{j}_{ai}\bar{Z}_{j})\\
=&\partial_{\mu}J_{a}^{\mu}+\frac{i}{2}(\partial^{\mu}\bar{Z}_{i}\Gamma^{i}_{aj}\partial_{\mu}\varphi^{j}+\partial^{\mu}Z^{i}\bar{\Gamma}^{j}_{ai}\partial_{\mu}\bar{\varphi}_{j}+\partial^{\mu}\bar{\varphi}_{i}T^{i}_{aj}\partial_{\mu}Z^{j}+\partial^{\mu}\varphi^{i}\bar{T}^{j}_{ai}\partial_{\mu}\bar{Z}_{j}\\
&-\frac{1}{2}\alpha \bar{Z}_{i}\Gamma^{i}_{aj}\varphi^{j}-\frac{1}{2}\alpha Z^{i}\bar{\Gamma}_{ai}^{j}\bar{\varphi}_{j}-\frac{1}{2}\alpha \bar{\varphi}_{i}T^{i}_{aj}Z^{j}-\frac{1}{2}\alpha\varphi^{i}\bar{T}^{j}_{ai}\bar{Z}_{j}+\frac{1}{2}\alpha\bar{\varphi}_{i}\Gamma^{i}_{aj}\varphi^{j}\\
&+\frac{1}{2}\alpha\varphi^{i}\bar{\Gamma}^{j}_{ai}\bar{\varphi}_{j}+\frac{1}{2}\alpha\bar{Z}_{i}T^{i}_{aj}Z^{j}+\frac{1}{2}\alpha Z^{i}\bar{T}^{j}_{ai}\bar{Z}_{j})-iM^{2}\bar{\varphi}_{i}\Gamma^{i}_{aj}\varphi^{j}-iM^{2}\varphi^{i}\bar{\Gamma}^{j}_{ai}\bar{\varphi}_{j}
\end{aligned}
\end{equation}
here the presence of $i$ emerges from the fact that the "conjugated" between $\varphi^{i}$ and $\bar{\varphi}_{i}$. The matter currents $J_{a}^{\mu}$ in present discussion are given by
\begin{equation}
\begin{aligned}
&J_{a}^{\mu}=-\frac{i}{2}(\partial^{\mu}\bar{Z}_{i}\Gamma^{i}_{aj}\varphi^{j}+\partial^{\mu}Z^{i}\bar{\Gamma}^{j}_{ai}\bar{\varphi}_{j}+\partial^{\mu}\bar{\varphi}_{i}T^{i}_{aj}Z^{j}+\partial^{\mu}\varphi^{i}\bar{T}^{j}_{ai}\bar{Z}_{j})
\end{aligned}
\end{equation}
a similar analysis of the local cohomology for Koszul-Tate differential $\delta$ at antighost number zero can be carried out for (3.7), we must ensure that the right-hand side of this equation be a total derivative term and in view of this, the following relations should hold
\begin{equation}
\begin{aligned}
&\Gamma_{aj}^{i}+\bar{T}_{aj}^{i}=0,\quad \quad \bar{\Gamma}_{aj}^{i}+T_{aj}^{i}=0,\quad \quad \Gamma_{aj}^{i}+\bar{\Gamma}_{aj}^{i}=0,\quad \quad T_{aj}^{i}+\bar{T}_{aj}^{i}=0
\end{aligned}
\end{equation}
while in the matrix form $(\Gamma_{a})_{ij}=\Gamma_{aj}^{i}$ we have
\begin{equation}
\begin{aligned}
\bar{\Gamma}_{a}=-\Gamma_{a},\quad \quad T_{a}=\Gamma_{a},\quad \quad \bar{T}_{a}=-\Gamma_{a}
\end{aligned}
\end{equation}
by using this, one can show that
 the conserved currents $J^{\mu}_{a}$ (3.8) can be obtained from the global invariance
\begin{equation}
\begin{aligned}
\Delta\varphi=i\xi^{a}\Gamma_{a}\varphi,\quad \quad\Delta\bar{\varphi}=-i\xi^{a}\Gamma_{a}^{T}\bar{\varphi},\quad \quad \Delta Z=i\xi^{a}\Gamma_{a}Z,\quad \quad\Delta\bar{Z}=-i\xi^{a}\Gamma_{a}^{T}\bar{Z}
\end{aligned}
\end{equation}
of the matter fields for arbitrary constants $\xi^{a}$ and the first-order deformation of the couplings part in complex scalar case takes the form of
\begin{equation}
\begin{aligned}
S_{1}^{m}&=-\frac{i}{2}\int d^{4}x((\partial^{\mu}\bar{Z})^{T}A_{\mu}\varphi-(\partial^{\mu}Z)^{T}A_{\mu}^{T}\bar{\varphi}+(\partial^{\mu}\bar{\varphi})^{T}A_{\mu}Z-(\partial^{\mu}\varphi)^{T}A_{\mu}^{T}\bar{Z})
\end{aligned}
\end{equation}

Thus it is significance to claim that the derivation of the second-order deformation of the massive complex scalar fields is apparently parallel to that
of the real scalar case by comparing (2.67) with (3.12).  We firstly compute $(S_{1}^{m},S_{1}^{A})$ and in the current situation, the procedure shown in the previous section demonstrates that
\begin{equation}
\begin{aligned}
S_{1}^{A}=\mathrm{tr}\int d^{4}x&(-iA^{*\mu}\left[\eta,A_{\mu}\right]-iB^{*\mu}\left[\eta,B_{\mu}\right]-i\varphi^{*T}\eta\varphi+i\bar{\varphi}^{*T}\eta^{T}\bar{\varphi}\\
&-iZ^{*T}\eta Z+i\bar{Z}^{*T}\eta^{T} \bar{Z}+\frac{i}{2}\eta^{*}\left[\eta,\eta\right])
 \end{aligned}
\end{equation}
as we discussed before, utilizing the canonical relations between fields and antifields we have
\begin{equation}
\begin{aligned}
&(\int d^{4}xJ_{a}^{\mu}A_{\mu}^{a},-\mathrm{tr}\int d^{4}yiA^{*\mu}\left[\eta,A_{\mu}\right])\\
=&\frac{1}{2}\int d^{4}x((\partial^{\mu}\bar{Z})^{T}\left[\eta,A_{\mu}\right]\varphi-(\partial^{\mu}Z)^{T}\left[\eta,A_{\mu}\right]^{T}\bar{\varphi}+(\partial^{\mu}\bar{\varphi})^{T}\left[\eta,A_{\mu}\right]Z-(\partial^{\mu}\varphi)^{T}\left[\eta,A_{\mu}\right]^{T}\bar{Z})
\end{aligned}
\end{equation}
together with
\begin{equation}
\begin{aligned}
&(\int d^{4}xJ_{a}^{\mu}A_{\mu}^{a},-\int d^{4}yi(\varphi^{*T}\eta\varphi-\bar{\varphi}^{*T}\eta^{T}\bar{\varphi}+Z^{*T}\eta Z-\bar{Z}^{*T}\eta^{T}\bar{Z}))\\
=&\frac{1}{2}\int d^{4}x(\bar{Z}^{T}\eta\partial^{\mu}(A_{\mu}\varphi)+(\partial^{\mu}\bar{Z})^{T}A_{\mu}\eta\varphi+(\eta Z)^{T}\partial^{\mu}(A_{\mu}^{T}\bar{\varphi})+(\partial^{\mu}Z)^{T}A_{\mu}^{T}\eta^{T}\bar{\varphi}\\
&+\bar{\varphi}^{T}\eta\partial^{\mu}(A_{\mu}Z)+(\partial^{\mu}\bar{\varphi})^{T}A_{\mu}\eta Z+(\eta \varphi)^{T}\partial^{\mu}(A_{\mu}^{T}\bar{Z})+(\partial^{\mu}\varphi)^{T}A_{\mu}^{T}\eta^{T}\bar{Z})
\end{aligned}
\end{equation}
combining it with (3.14) we obtain
\begin{equation}
\begin{aligned}
&(S_{1}^{m}+S_{1}^{A},S_{1}^{m}+S_{1}^{A})\\
=&\int d^{4}x(\bar{Z}^{T}\eta\partial^{\mu}(A_{\mu}\varphi)+(\partial^{\mu}\bar{Z})^{T}\eta A_{\mu}\varphi+(\eta Z)^{T}\partial^{\mu}(A_{\mu}^{T}\bar{\varphi})+(\partial^{\mu}Z)^{T}\eta ^{T}A_{\mu}^{T}\bar{\varphi}\\
&+\bar{\varphi}^{T}\eta\partial^{\mu}(A_{\mu}Z)+(\partial^{\mu}\bar{\varphi})^{T}\eta A_{\mu}Z+(\eta \varphi)^{T}\partial^{\mu}(A_{\mu}^{T}\bar{Z})+(\partial^{\mu}\varphi)^{T}\eta^{T}A_{\mu}^{T}\bar{Z})
\end{aligned}
\end{equation}
on the other hand, similar to the real scalar fields let us investigate
\begin{equation}
\begin{aligned}
&s(\int d^{4}x((A^{\mu}\varphi)^{T}A^{T}_{\mu}\bar{Z}+(A^{\mu}Z)^{T}A^{T}_{\mu}\bar{\varphi}))\\
=&\int d^{4}x((\partial^{\mu}\eta\varphi)^{T}A^{T}_{\mu}\bar{Z}+(A^{\mu}\varphi)^{T}(\partial_{\mu}\eta)^{T}\bar{Z}+(\partial^{\mu}\eta Z)^{T}A^{T}_{\mu}\bar{\varphi}+(A^{\mu}Z)^{T}(\partial_{\mu}\eta)^{T}\bar{\varphi})\\
=&-\int d^{4}x((\eta\partial^{\mu}\varphi)^{T} A_{\mu}^{T}\bar{Z}+(\eta\varphi)^{T}\partial^{\mu}(A_{\mu}^{T}\bar{Z})+\partial_{\mu}(A^{\mu}\varphi)^{T}\eta^{T}\bar{Z}+(A^{\mu}\varphi)^{T}\eta^{T}\partial_{\mu}\bar{Z}\\
&+(\eta\partial^{\mu}Z)^{T} A_{\mu}^{T}\bar{\varphi}+(\eta Z)^{T}\partial^{\mu}(A_{\mu}^{T}\bar{\varphi})+\partial_{\mu}(A^{\mu}Z)^{T}\eta^{T}\bar{\varphi}+(A^{\mu}Z)^{T}\eta^{T}\partial_{\mu}\bar{\varphi})
\end{aligned}
\end{equation}
an immediate outcome of the above results is the determination of $S_{2}^{m}$ which can be directly seen upon using the fact
\begin{equation}
\begin{aligned}
\partial_{\mu}(A^{\mu}\varphi)^{T}\eta^{T}\bar{Z}=(\partial_{\mu}(A^{\mu}\varphi)^{T}\eta^{T}\bar{Z})^{T}=\bar{Z}^{T}\eta\partial^{\mu}(A_{\mu}\varphi)
\end{aligned}
\end{equation}
analogously for $(\eta\partial^{\mu}Z)^{T}G$ and we assert that the equation (2.68) is solved by taking
\begin{equation}
\begin{aligned}
S_{2}^{m}=&\frac{1}{2}\int d^{4}x\left[(A^{\mu}\varphi)^{T}A^{T}_{\mu}\bar{Z}+(A^{\mu}Z)^{T}A^{T}_{\mu}\bar{\varphi}\right]\\
\end{aligned}
\end{equation}

The same process may be followed if we search for the third-order deformation $S_{3}^{m}$ for the couplings part and we should calculate
\begin{equation}
\begin{aligned}
&(S_{2}^{m},-\mathrm{tr}\int d^{4}yiA^{*\mu}\left[\eta,A_{\mu}\right])\\
=&-\frac{i}{2}\int d^{4}x((\left[\eta,A^{\mu}\right]\varphi)^{T}A^{T}_{\mu}\bar{Z}+(A^{\mu}\varphi)^{T}\left[\eta,A_{\mu}\right]^{T}\bar{Z}+(\left[\eta,A^{\mu}\right]Z)^{T}A^{T}_{\mu}\bar{\varphi}\\
&+(A^{\mu}Z)^{T}\left[\eta,A_{\mu}\right]^{T}\bar{\varphi})
\end{aligned}
\end{equation}
as well as
\begin{equation}
\begin{aligned}
&(S_{2}^{m},-\int d^{4}iy(\varphi^{*T}\eta\varphi-\bar{\varphi}^{*T}\eta^{T}\bar{\varphi}+Z^{*T}\eta Z-\bar{Z}^{*T}\eta^{T} \bar{Z}))\\
=&-\frac{i}{2}\int d^{4}x((A^{\mu}\eta\varphi)^{T}A^{T}_{\mu}\bar{Z}-(A^{\mu}\varphi)^{T}A^{T}_{\mu}\eta^{T}\bar{Z}-(A^{\mu}Z)^{T}A^{T}_{\mu}\eta^{T}\bar{\varphi}+(A^{\mu}\eta Z)^{T}A^{T}_{\mu}\bar{\varphi})
\end{aligned}
\end{equation}
expanding $\left[\quad ,\quad \right]$ in (3.20) and a simple algebraic manipulation shows that
\begin{equation}
\begin{aligned}
(S_{1},S_{2}^{m})=0
\end{aligned}
\end{equation}
which means $S_{i}=0$ for $i\geq3$ as we exhibit before. Now following the lines discussed in the real scalar case, the antighost number zero part $\bar{S}_{0}\left[A_{\mu},\varphi,\bar{\varphi},Z,\bar{Z}\right]$ of the deformed master action $S$ can be written in a compact formula
\begin{equation}
\begin{aligned}
\bar{S}_{0}=&\mathrm{tr}\int d^{4}x(-\frac{1}{4}\mathcal{F}_{\mu\nu}\mathcal{F}^{\mu\nu}-\frac{m^{2}}{2}B_{\mu}B^{\mu}+m^{2}D_{\mu}B_{\nu}\mathcal{F}^{\mu\nu}+\frac{1}{2}((d_{\mu}Z)^{T} \bar{d}^{\mu}\bar{\varphi}\\
&+(d^{\mu}\varphi)^{T}\bar{d}_{\mu}\bar{Z})+\frac{1}{4}\alpha\varphi^{T}\bar{\varphi}+\frac{1}{4}\alpha Z^{T}\bar{Z}-\frac{1}{4}\alpha(\varphi^{T}\bar{Z}+\bar{\varphi}^{T}Z)-M^{2}\bar{\varphi}^{T}\varphi)
\end{aligned}
\end{equation}
here we use the notation
\begin{equation}
\begin{aligned}
 d_{\mu}=\partial_{\mu}-igA_{\mu},\quad \quad  \bar{d}^{\mu}=\partial^{\mu}+igA^{\mu T}
\end{aligned}
\end{equation}
and taking advantage of the equations of motion of auxiliary fields
\begin{equation}
\begin{aligned}
Z=\varphi+\frac{2}{\alpha}d_{\mu}d^{\mu}\varphi,\quad \quad \quad \bar{Z}=\bar{\varphi}+\frac{2}{\alpha}\bar{d}_{\mu}\bar{d}^{\mu}\bar{\varphi}
\end{aligned}
\end{equation}
we can rewrite the action (3.23) in the equivalent form of
\begin{equation}
\begin{aligned}
\bar{S}_{0}=\mathrm{tr}\int d^{4}x(&-\frac{1}{4}\mathcal{F}_{\mu\nu}\mathcal{F}^{\mu\nu}+\frac{m^{2}}{2}D_{\mu}\mathcal{F}^{\mu\nu}D^{\lambda}\mathcal{F}_{\lambda\nu}+(d_{\mu}\varphi)^{T}\bar{d}^{\mu}\bar{\varphi}\\
&-\frac{1}{\alpha}(d_{\mu}d^{\mu}\varphi)^{T}\bar{d}^{\mu}\bar{d}_{\mu}\bar{\varphi}-M^{2}\bar{\varphi}^{T}\varphi)\\
\end{aligned}
\end{equation}
here $\mathcal{F}_{\mu\nu}=F_{\mu\nu}-ig\left[A_{\mu},A_{\nu}\right],D_{\mu}=\partial_{\mu}-ig\left[A_{\mu},A_{\nu}\right]$ and the main differences compared to the real scalar case are caused by the extra factor $i$, furthermore the local gauge transformations after the deformation procedure are modified as
\begin{equation}
\begin{aligned}
\delta_{\lambda}A_{\mu}^{a}=\partial_{\mu}\lambda^{a}-igf^{a}_{bc}A_{\mu}^{b}\lambda^{c},\quad \quad \quad \delta_{\lambda}\varphi^{i}=ig\Gamma^{i}_{aj}\varphi^{j}\lambda^{a},\quad \quad \quad \delta_{\lambda}\bar{\varphi}_{i}=-ig\Gamma^{j}_{ai}\bar{\varphi}_{j}\lambda^{a}
\end{aligned}
\end{equation}
\section{Massive Dirac spinor fields}
We shall consider the following free Lagrangian density between the Abelian gauge fields $A_{\mu}$ and the Dirac spinor fields $(\psi^{\alpha}_{i},\bar{\psi}_{\alpha}^{i})$ with higher derivative terms described by
\begin{equation}
\begin{aligned}
\mathcal{L}=\mathcal{L}_{GE}+\bar{\psi}_{\alpha}^{i}(\gamma^{\mu}\partial_{\mu})^{\alpha}_{\beta}\psi^{\beta}_{i}+(\partial_{\omega}\bar{\psi}_{\sigma}^{i}\gamma^{\omega\sigma}_{\alpha})((\gamma^{\nu}\partial_{\nu})^{\alpha}_{\tau}(\gamma^{\mu}\partial_{\mu})^{\tau}_{\beta})\psi^{\beta}_{i}-M\bar{\psi}_{\alpha}^{i}\psi^{\alpha}_{i}
\end{aligned}
\end{equation}
here $\mathcal{L}_{GE}$ is the generalized electrodynamics density and $\gamma^{\mu}$ are the standard Dirac's gamma matrices satisfying $\{\gamma^{\mu},\gamma^{\nu}\}=2\delta^{\mu\nu}$. The equations of motion of the Dirac fields in the component form are
\begin{equation}
\begin{aligned}
((\gamma^{\mu}\partial_{\mu})^{\sigma}_{\beta}-(\gamma^{\omega}\partial_{\omega})^{\sigma}_{\alpha}(\gamma^{\nu}\partial_{\nu})^{\alpha}_{\tau}(\gamma^{\mu}\partial_{\mu})^{\tau}_{\beta})\psi^{\beta}_{i}-M\psi^{\sigma}_{i}=0
\end{aligned}
\end{equation}
instead of the Lagrangian (4.1), for convenience we will adopt the matrix form $\psi=(\psi_{1},...,\psi_{M})^{T}$ and $\psi_{i}=(\psi^{\alpha}_{i})$ together with $\bar{\psi}=(\bar{\psi}^{1},...,\bar{\psi}^{M}),(\bar{\psi}^{i})=(\bar{\psi}_{\alpha}^{i})$ and by means of the auxiliary Dirac fields $Z=(Z^{\alpha}_{i})^{T},\bar{Z}=(\bar{Z}_{\alpha}^{i})$, the above Lagrangian density is converted into the following equivalent one
\begin{equation}
\begin{aligned}
\mathcal{\tilde{L}}=&-\frac{1}{4}F_{\mu\nu}^{a}F^{\mu\nu}_{a}-\frac{m^{2}}{2}B^{a}_{\mu}B^{\mu }_{a}+m^{2}\partial_{\mu}B_{\nu}^{a}F^{\mu\nu}_{a}-(\partial_{\mu}\bar{\psi}^{i}\gamma^{\mu})(\gamma^{\nu}\partial_{\nu}) Z_{i}\\
&+\bar{Z}^{i}(\gamma^{\mu}\partial_{\mu})\psi_{i}+\bar{\psi}^{i}(\gamma^{\mu}\partial_{\mu})\psi_{i}+\bar{Z}^{i}Z_{i}-M\bar{\psi}^{i}\psi_{i}
\end{aligned}
\end{equation}
written explicitly, the equations of motion for $(\psi,\bar{\psi})$ and the auxiliary fields $(Z,\bar{Z})$ are governed by
\begin{equation}
\begin{aligned}
&(\gamma^{\mu}\partial_{\mu})\psi_{i}+Z_{i}=0,\quad \quad (\gamma^{\mu}\partial_{\mu})(\gamma^{\nu}\partial_{\nu})Z_{i}+(\gamma^{\mu}\partial_{\mu})\psi_{i}-M\psi_{i}=0,\\
&(\partial_{\mu}\partial_{\nu}\bar{\psi}^{i})\gamma^{\mu}\gamma^{\nu}+\bar{Z}^{i}=0,\quad \quad \partial_{\mu}\bar{Z}^{i}\gamma^{\mu}+\partial_{\mu}\bar{\psi}^{i}\gamma^{\mu}+M\bar{\psi}^{i}=0
\end{aligned}
\end{equation}
plugging these equations into (4.3), it is easy to return to the original Lagrangian density as well as the dynamic equations (4.2). By analogy with the previous discussions, the local gauge transformations are
\begin{equation}
\begin{aligned}
\Delta_{\lambda} A_{\mu}^{a}=\partial_{\mu}\lambda^{a},\quad  \Delta_{\lambda} \psi^{\alpha}_{i}=0,\quad \Delta_{\lambda} \bar{\psi}_{\alpha}^{i}=0,\quad \Delta_{\lambda} Z^{\alpha}_{i}=0,\quad \Delta_{\lambda}  \bar{Z}_{\alpha}^{i}=0
\end{aligned}
\end{equation}

In the case of the Dirac spinor fields, the non-trivial action of Koszul-Tate differential on the generators $\psi^{*i},\bar{\psi}^{*}_{i},Z^{*i},\bar{Z}^{*}_{i}$ of the BRST complex present in the following way
\begin{equation}
\begin{aligned}
& \delta\psi^{*i}_{\alpha}=(\partial_{\mu}\bar{Z}^{i}\gamma^{\mu})_{\alpha}+(\partial_{\mu}\bar{\psi}^{i}\gamma^{\mu})_{\alpha}+M\bar{\psi}^{i}_{\alpha},\quad \quad \quad \quad \delta Z^{*i}_{\alpha}=-(\partial_{\mu}\partial_{\nu}\bar{\psi}^{i}\gamma^{\mu}\gamma^{\nu})_{\alpha}-\bar{Z}^{i}_{\alpha},\\ &\delta\bar{\psi}^{*\alpha}_{i}=-(\gamma^{\mu}\partial_{\mu}\gamma^{\nu}\partial_{\nu}Z_{i})^{\alpha}-(\gamma^{\mu}\partial_{\mu}\psi_{i})^{\alpha}+M\psi_{i}^{\alpha},\quad \delta \bar{Z}_{i}^{*\alpha}=-(\gamma^{\mu}\partial_{\mu}\psi_{i})^{\alpha}-Z_{i}^{\alpha}
\end{aligned}
\end{equation}
applying these and considering the ansatz as shown in (2.39) we get
\begin{equation}
\begin{aligned}
&\delta(-i\psi^{*i}\Gamma^{j}_{ai}\psi_{j}-i\bar{\psi}^{i}\bar{\Gamma}_{ai}^{j}\bar{\psi}_{j}^{*}-iZ^{*i}T^{j}_{ai}Z_{j}-i\bar{Z}^{i}\bar{T}^{j}_{ai}\bar{Z}_{j}^{*})\\
=&\partial_{\mu}J_{a}^{\mu}+i\bar{Z}^{i}\gamma^{\mu}\Gamma^{j}_{ai}\partial_{\mu}\psi_{j}+i\bar{\psi}^{i}\gamma^{\mu}\Gamma^{j}_{ai}\partial_{\mu}\psi_{j}-iM\bar{\psi}^{i}\Gamma^{j}_{ai}\psi_{j}-i(\partial_{\mu}\bar{\psi}^{i}\gamma^{\mu})\bar{\Gamma}_{ai}^{j}(\gamma^{\nu}\partial_{\nu})Z_{j}\\
&+i\bar{\psi}^{i}\bar{\Gamma}_{ai}^{j}\gamma^{\mu}\partial_{\mu}\psi_{j}-iM\bar{\psi}^{i}\bar{\Gamma}_{ai}^{j}\psi_{j}-i(\partial_{\nu}\bar{\psi}^{i})\gamma^{\mu}\gamma^{\nu}T^{j}_{ai}\partial_{\mu}Z_{j}+i\bar{Z}^{i}T^{j}_{ai}Z_{j}\\
&+i\bar{Z}^{i}\bar{T}^{j}_{ai}(\gamma^{\mu}\partial_{\mu})\psi_{j}+i\bar{Z}^{i}\bar{T}^{j}_{ai}Z_{j}
\end{aligned}
\end{equation}
here the matter currents are given by
\begin{equation}
\begin{aligned}
&J^{\mu}_{a}=i((\partial_{\nu}\bar{\psi}^{i}\gamma^{\nu})\gamma^{\mu}T^{j}_{ai}Z_{j}+(\bar{\psi}^{i}\gamma^{\mu})\bar{\Gamma}_{ai}^{j}(\gamma^{\nu}\partial_{\nu}Z_{j})-\bar{Z}^{i}\Gamma^{j}_{ai}\gamma^{\mu}\psi_{j}-\bar{\psi}^{i}\Gamma_{ai}^{j}\gamma^{\mu}\psi_{j})
\end{aligned}
\end{equation}
 imposition of the non-trivial local cohomology for $\delta$ will result in
\begin{equation}
\begin{aligned}
\Gamma^{j}_{ai}+\bar{T}^{j}_{ai}=0,\quad  \Gamma^{j}_{ai}+\bar{\Gamma}_{ai}^{j}=0,\quad\ \bar{\Gamma}_{ai}^{j}+T^{j}_{ai}=0,\quad T^{j}_{ai}+\bar{T}^{j}_{ai}=0
\end{aligned}
\end{equation}
when expressed in matrix form these relations read
\begin{equation}
\begin{aligned}
T_{a}=\Gamma_{a},\quad\quad   \bar{T}_{a}=-\Gamma_{a},\quad \quad \bar{\Gamma}_{a}=-\Gamma_{a}
\end{aligned}
\end{equation}
of course, also it is not hard to deduce the conserved currents $J^{\mu}_{a}$ from the symmetry transformations
\begin{equation}
\begin{aligned}
\Delta\psi^{\alpha}=i\xi^{a}\Gamma_{a}\psi^{\alpha},\quad \Delta\bar{\psi}_{\alpha}=-i\xi^{a}\Gamma_{a}^{T}\bar{\psi}_{\alpha},\quad \Delta Z^{\alpha}=i\xi^{a}\Gamma_{a}Z^{\alpha},\quad \Delta\bar{Z}_{\alpha}=-i\xi^{a}\Gamma_{a}^{T}\bar{Z}_{\alpha}
\end{aligned}
\end{equation}
of the spinor fields for the global constants $\xi^{a}$. Then setting $A_{\mu}=A_{\mu}^{a}\Gamma_{a}$ and this form of the gauge fields suggests that we can construct the first-order deformation of the spinor fields as
\begin{equation}
\begin{aligned}
S_{1}^{m}=\int d^{4}xi((\partial_{\nu}\bar{\psi})^{T}\gamma^{\nu}\gamma^{\mu}A_{\mu}Z-(\bar{\psi}\gamma^{\mu})^{T}A_{\mu}(\gamma^{\nu}\partial_{\nu}Z)-\bar{Z}^{T}A_{\mu}\gamma^{\mu}\psi-\bar{\psi}^{T}A_{\mu}\gamma^{\mu}\psi)
\end{aligned}
\end{equation}
it is worth commenting that in the above expression the matrices $\Gamma_{a}$ act upon the components in the matter field vectors while the Dirac's gamma matrices operate on the components of spinors, these two actions are independent and commute with each other, hence we could exchange them from left to right or the other way round safely.

Now the basic problem is again to seek for the solutions of (2.68) and as can be seen from the above results we acquire
\begin{equation}
\begin{aligned}
S_{1}^{A}=\mathrm{tr}\int d^{4}x&i(-A^{*\mu}\left[\eta,A_{\mu}\right]-B^{*\mu}\left[\eta,B_{\mu}\right]-\psi^{*T}\eta\psi+\bar{\psi}^{T}\eta^{T}\bar{\psi}^{*}\\
&-Z^{*T}\eta Z+\bar{Z}^{T}\eta^{T}\bar{Z}^{*}+\frac{1}{2}\eta^{*}\left[\eta,\eta\right])
 \end{aligned}
\end{equation}
a straightforward calculation thus yields
\begin{equation}
\begin{aligned}
(S_{1}^{m}+S_{1}^{A},S_{1}^{m}+S_{1}^{A})=&-2\int d^{4}x((\partial_{\nu}\bar{\psi})^{T}\gamma^{\nu}\gamma^{\mu}\eta A_{\mu} Z+\bar{\psi}^{T}\eta\gamma^{\nu}\gamma^{\mu}\partial_{\nu}(A_{\mu}Z)\\
&+\bar{\psi}^{T}\gamma^{\mu}A_{\mu}\eta\gamma^{\nu}\partial_{\nu}Z+\partial_{\nu}(\bar{\psi}^{T}A_{\mu})\gamma^{\mu}\gamma^{\nu}\eta Z)
\end{aligned}
\end{equation}
and as a matter of fact, it is easy to check that the solution is expressible in the form of
\begin{equation}
\begin{aligned}
S_{2}^{m}=-\int d^{4}x\bar{\psi}^{T}\gamma^{\nu}\gamma^{\mu}A_{\nu}A_{\mu}Z
\end{aligned}
\end{equation}
then we have to go further to construct the third-order deformation and indeed we find
\begin{equation}
\begin{aligned}
(S_{1},S_{2}^{m})=0
\end{aligned}
\end{equation}
which is not hard to examine and hence we conclude that $S_{i}=0$ for $i\geq3$. In the same manner, the antighost number zero part $\bar{S}_{0}\left[A_{\mu},\psi,\bar{\psi},Z,\bar{Z}\right]$ of the solution of the deformation master equations takes the expression as
\begin{equation}
\begin{aligned}
\bar{S}_{0}=\mathrm{tr}\int d^{4}x(&-\frac{1}{4}\mathcal{F}_{\mu\nu}\mathcal{F}^{\mu\nu}-\frac{m^{2}}{2}B_{\mu}B^{\mu}+m^{2}D_{\mu}B_{\nu}\mathcal{F}^{\mu\nu}-(\bar{d}_{\mu}\bar{\psi})^{T}\gamma^{\mu}\gamma^{\nu}d_{\nu} Z\\
&+\bar{Z}^{T}\gamma^{\mu}d_{\mu}\psi+\bar{\psi}^{T}\gamma^{\mu}d_{\mu}\psi+\bar{Z}^{T}Z-M\bar{\psi}^{T}\psi)
\end{aligned}
\end{equation}
 then making using of the equations of motion of auxiliary spinor fields
\begin{equation}
\begin{aligned}
Z=-\gamma^{\mu}d_{\mu}\psi,\quad \quad \quad \bar{Z}=-\bar{d}_{\nu}\bar{d}_{\mu}\bar{\psi}\gamma^{\mu}\gamma^{\nu}
\end{aligned}
\end{equation}
we arrive at the following equivalent form of the Lagrangian in terms of the classical dynamic fields
\begin{equation}
\begin{aligned}
\bar{S}_{0}=\mathrm{tr}\int d^{4}x(&-\frac{1}{4}\mathcal{F}_{\mu\nu}\mathcal{F}^{\mu\nu}+\frac{m^{2}}{2}D_{\mu}\mathcal{F}^{\mu\nu}D^{\lambda}\mathcal{F}_{\lambda\nu}+\bar{\psi}^{T}\gamma^{\mu}d_{\mu}\psi-M\bar{\psi}^{T}\psi\\
&+(\bar{d}_{\mu}\bar{\psi})^{T}\gamma^{\mu}\gamma^{\nu}d_{\nu}\gamma^{\omega}d_{\omega}\psi)
\end{aligned}
\end{equation}
apparently, such Lagrangian density describes the interactions among the Abelian gauge fields and a collection of massive spinor fields with higher order derivative terms and we emphasize again that the deformed system is invariant under the following non-Abelian gauge transformations
\begin{equation}
\begin{aligned}
\delta_{\lambda}A_{\mu}^{a}=\partial_{\mu}\lambda^{a}-igf^{a}_{bc}A_{\mu}^{b}\lambda^{c},\quad \quad \quad \delta_{\lambda}\psi^{\alpha}_{i}=-i\Gamma^{j}_{ai}\psi^{\alpha}_{j}\lambda^{a},\quad\quad \quad \delta_{\lambda}\bar{\psi}_{\alpha}^{i}=i\Gamma^{i}_{aj}\bar{\psi}_{\alpha}^{j}\lambda^{a}
\end{aligned}
\end{equation}
\section{Conclusion and discussion}
In this paper, we primarily propose the derivation of the consistent interactions in Yang-Mills gauge theory coupled to matter fields with higher derivative terms. We make using of the auxiliary fields to reduce the higher derivative to first-order and the number of dynamical variables of the resulting system will be twice than the original one. Then applying the standard BRST deformations procedure and utilizing the equations of motion of the auxiliary fields, we are capable of obtaining the added interaction terms which are consistent with the deformation master equations. A natural generalization of this work is the construction of consistent interactions from the Hamiltonian BRST-invariant deformation approach in such higher derivative free theories. More precisely, in the reduced formalism, it is convenient to establish the BRST charge as well as the BRST-invariant Hamiltonian from the standard method in the procedure of BRST quantization and in order to derive the consistent interactions, we express the deformed Hamiltonian and the BRST charge in terms of power series expansion of the deformation parameter. Afterwards, it is reasonable to require that the nilpotency of the BRST charge and the commutativity between the Hamiltonian and BRST charge should be preserved which will lead to a set of iterative equations coming from the perturbative expansion order by order. Through solving these equations recursively we are able to gain various consistent interactions at different orders in the original system both for gauge and matter fields or among a collection of gauge fields. We might obtain the Abelian and the non-Abelian Lagrangian action by extracting the first-class Hamiltonian of the interacting theory after the deformation process as we can imagine. Another useful method to deal with the higher-order derivative field theories is the Ostrogradsky formalism and in this approach, we enlarge the number of the canonical momenta from the higher-order time derivatives of the dynamic variables. In such extended phase space, the Hamiltonian of the higher derivative system can be cast in terms of these additional canonical phase coordinates in a first-order form which is more familiar and tractable. Following the lines of the standard Hamiltonian BRST deformations, it makes possible to deduce the consistent interactions and we speculate that these deformation terms will be identified with those derived in the order reduction method mentioned above. All of these would be interesting to exploit in future.

\appendix

\acknowledgments

 The author would like to thank G.W.Wan for long time encouragements and is grateful to S.M.Zhu for useful support.


\end{document}